*Topical Review*

**Quantum Anomalous Hall Effect in Time-Reversal-Symmetry Breaking Topological Insulators**

Cui-Zu Chang*[1], Mingda Li[2,3]

[1]Francis Bitter Magnet Lab, Massachusetts Institute of Technology, Cambridge, MA 02139, USA

[2]Department of Mechanical Engineering, Massachusetts Institute of Technology, Cambridge, MA 02139, USA

[3]Department of Nuclear Engineering, Massachusetts Institute of Technology, Cambridge, MA 02139, USA

*Email: czchang@mit.edu

**Abstract: The quantum anomalous Hall effect (QAHE), the last member of Hall family, was predicted to exhibit quantized Hall conductivity $\sigma_{yx} = \dfrac{e^2}{h}$ without any external magnetic field. The QAHE shares a similar physical phenomenon with the integer quantum Hall effect (QHE), whereas its physical origin relies on the intrinsic topological inverted band structure and ferromagnetism. Since the QAHE does not require external energy input in the form of magnetic field, it is believed that this effect has unique potential for applications in future electronic devices with low-power consumption. More recently, the QAHE has been experimentally observed in thin films of the time-reversal symmetry breaking ferromagnetic (FM) topological insulators (TI), Cr- and V- doped $(Bi,Sb)_2Te_3$. In this *Topical Review*, we review the history of TI based QAHE, the route to the experimental**

observation of the QAHE in the above two systems, the current status of the research of the QAHE, and finally the prospects for future studies.

**Content**

1. Introduction
   1.1 The Hall family
   1.2 Topological insulators (TIs) and magnetic TIs
2. Theoretical prediction of QAHE based on TI
   2.1 QAHE in magnetic two-dimensional TIs
   2.2 QAHE in TI/ferromagnetic insulator heterostructures
   2.3 QAHE in magnetic three-dimensional TIs
   2.4 The pathways to break the time-reversal symmetry in TIs
3. The route to the experimental observation of QAHE
   3.1 Quantum well TI films on insulating substrates
   3.2 Inducing FM order into the quantum well TI films
   3.3 Tuning the chemical potential in the surface states exchange gap
4. QAHE in Cr-doped TIs
   4.1 First experimental observation of the QAHE
   4.2 Trajectory of QAHE
   4.3 Nonlocal transport in QAHE
   4.4 Zero Hall plateau in QAHE
5. QAHE in V-doped TIs
   5.1 Experimental verification of van Vleck type ferromagnetism
   5.2 The advantages of V-doped TI system
   5.3 High-precision and self-driven QAHE
   5.4 The relationship between the impurity bands and QAHE
   5.5 Zero-field dissipationless chiral edge transport in  the QAHE
   5.6 The nature of dissipation in QAHE

## 1. Introduction

The quantum anomalous Hall effect (QAHE), the quantized version of the anomalous Hall effect (AHE), is generally believed to be the last member of the Hall family. After the discovery of integer quantum Hall effect (QHE) [1,2], theorists found that the QHE theory can also explain the AHE in ferromagnetic (FM) materials, where the Berry curvature used for explaining QHE can also determine the AH conductivity. However, the Hall resistance is not quantized in a normal FM metal, since the valence band is partially filled and thus the integral is not taken over the entire Brillouin zone [3]. In fact, if one can achieve a two-dimensional (2D) FM insulator with a non-zero Chern number, the QAHE can be realized. In reality, finding these FM insulators with a non-zero Chern number is a great challenge, thus for decades, little progress has been made in the experimental realization of QAHE.

The historic discovery of topological insulator (TI) materials beginning in 2006 made the search for a FM insulator with a non-zero Chern number, called the QAH insulator, become practical [4,5]. Through careful sample optimization and exploration, the QAHE was finally observed in FM TI thin films [6,7]. The experimental observation of QAHE concludes a decades-long search for the QHE without Landau levels, first predicted by F. Duncan M. Haldane in 1988 [8].

In this *Topical Review*, we review the QAHE in time-reversal symmetry (TRS) -breaking TIs, with a focus on recent experimental progress. There are already an excellent review of QAHE from a theoretical perspective [9], and another excellent review mainly focusing on magnetic TIs but also mentioning Cr-based QAHE[10]. In Section 1, we briefly introduce the Hall family and intrinsic and magnetically-doped TI; in Section 2, we present the relevant theoretical prediction about the QAHE in TRS-breaking TI; in Section 3, the route leading to the experimental observation of the QAHE in TRS-breaking TI is discussed; in Section 4 and 5, we review the recent progress on the QAHE in Cr- and V-doped TI systems, respectively. We close with the summary and the prospects for future studies of QAHE in Section 6.

## 1.1 The Hall family

When electric current flows through a conducting material placed in a perpendicular external magnetic field, the charge carriers that constitute the current are pushed toward the side of the sample by Lorentz force. Therefore, a transverse voltage, named Hall voltage, is developed across the sample (Fig. 1a). This effect, called the Hall effect, was discovered by E. H. Hall in 1879 while he was a student [11]. The Hall effect has been widely used to identify the carrier type, and measure the density and mobility of the carriers in conducting materials, and also provides a direct method to measure magnetic fields. Shortly after the discovery of the Hall effect, in 1881, E. H. Hall performed similar measurements on the FM materials nickel (Ni) and cobalt (Co). He found that the Hall resistance shows an unusually large slope at low magnetic fields [12]. The unusual Hall effect in FM materials was later called the AHE (Fig. 1b). In fact, contrary to the original ordinary Hall effect where the effect is proportional to external magnetic field, in AHE the Hall voltage is approximately proportional to the magnetization. The ordinary Hall effect and the AHE have been the only two members of the Hall family for about one century.

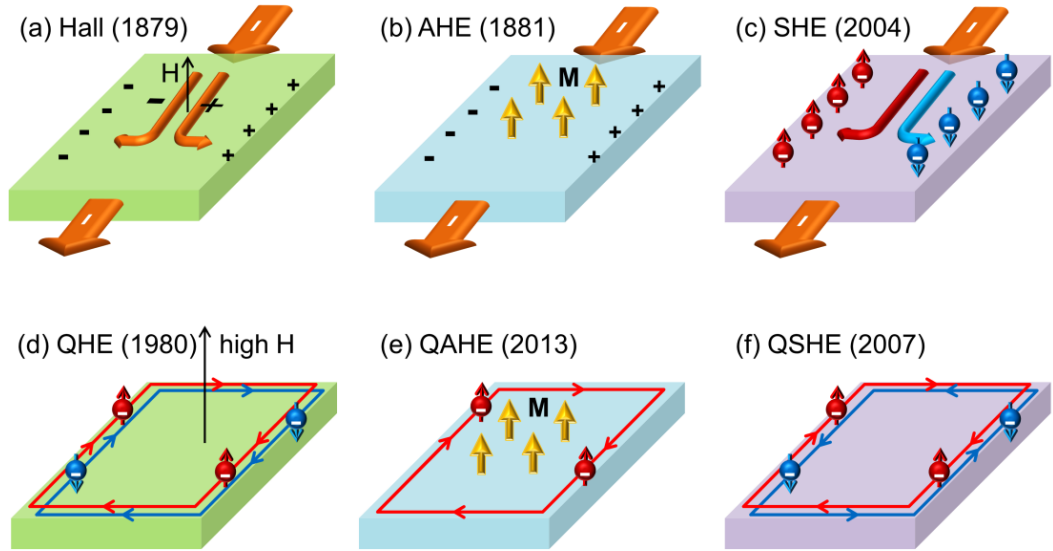

**Fig. 1| Members of Hall family.** (a) Hall effect. (b) AHE. (c) SHE. (d) QHE, (e) QAHE. (f) QSHE. Numbers in parentheses indicate the years of each discovery. $H$ is the external magnetic field, $M$ is the intrinsic spontaneous magnetization, and $S$ denotes spin.

In 1980, the integer QHE, the quantized version of the Hall effect, was discovered by von Klitzing in the Si/SiO$_2$ field-effect transistor in a strong external magnetic field (Fig. 1d) [1]. This discovery introduced the first quantized Hall member into the Hall family. In a strong external magnetic field, the continuous density of states of the two-dimensional electron gas (2DEG) split into equally-spaced Landau levels. When the Fermi level of the system lies between two neighboring Landau levels, the bulk carriers are localized, but the electrons can propagate along the edge of the sample. In this case, the Hall resistance forms well-defined plateaus and the longitudinal resistance ideally becomes exactly zero. Despite the great potential of the dissipationless edge states of the QHE for future low-power consumption and high-speed electronic devices, the necessity of strong magnetic fields and the consequent high-power input impedes the application of such edge states in QHE.

The spin Hall effect (SHE), another basic member of Hall family in addition to ordinary Hall effect and AHE, was first predicted by M. I. Dyakonov et al. in 1971 [13,14]. Unlike the ordinary Hall effect where the positive and negative charge carriers accumulate at opposite edges of the sample, in SHE, spin-up and spin-down carriers accumulate at the opposite edges of the sample (Fig. 1c). This effect was not observed until more than 30 years later in thin films of the *p*-type doped GaAs and InGaAs semiconductors using scanning Kerr rotation measurements in 2004 [15]. One reason that it took over 30 years to observe SHE since its prediction is that most of the early attempts to realize SHE were based on external perturbation such as spin injection or external magnetic field [16,17]. However, these approaches in fact only generate *extrinsic* SHE which involves anisotropic skew scattering off impurities and is too weak to support observable SHE; the observation, on the contrary, is based on a newer proposal using the Rashba spin-orbit split bands, which is an *intrinsic* mechanism.

One year after the experimental observation of the SHE, in 2005, the quantized version of the SHE, the so-called the quantum spin Hall effect (QSHE), was first theoretically proposed in graphene by C. Kane et al. [18]. The Kane QSHE model can be considered as two copies of the Haldane model, where the spin-up electrons exhibit a chiral QHE while the spin-down electrons exhibit an anti-chiral QHE (Fig. 1e). In reality, the QSHE was observed in the HgTe/CdTe quantum well structure by L. Molenkamp et al. in 2007 [5] after the prediction of S. C. Zhang et al. in 2006 [4]. The QSHE in HgTe/CdTe quantum wells can also be regarded as 2D TI, which will be discussed in detail later.

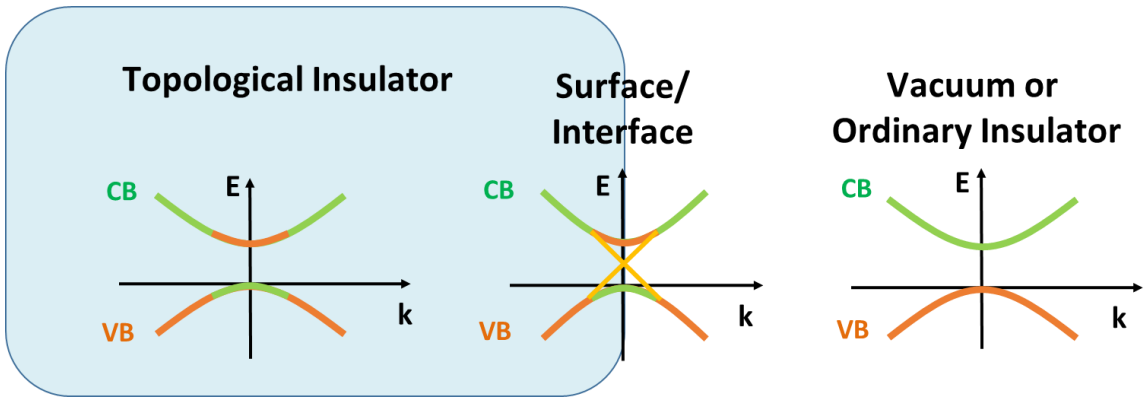

**Fig. 2| Illustration of a topological insulator (TI).** Compared with an ordinary insulator (right) where conduction band (CB, in green) and valance band (VB, in orange) is separated by an energy gap, the inversion between CB and VB in a TI (left) leads to conducting surface states (in yellow) at TI-ordinary insulator interfaces, including a TI surface with a vacuum.

As discussed above, the pioneering Haldane model proposed the QHE without an external magnetic field, aka QAHE. This model describes graphene in a periodic magnetic field with zero net magnetic flux and offers the possibility of realizing the QAHE. However, it did not provide any strategy for realizing it in practical systems. Eventually, the discovery of TIs provided a feasible platform for realizing the QAHE - the strong spin-orbit coupling not only provides the prerequisite to realize AHE, but also drives the inversion of the subbands of opposite spin under strong exchange field splitting, which finally enables the quantization of Hall conductance [5,19]. The QAHE was finally observed in thin films of magnetically doped TIs by Chang et al. in 2013 for the first time [6].

## 1.2 Topological insulators (TIs) and magnetic TIs

TIs have radically changed the research landscape of condensed matter physics since 2006 [18]. The first experimentally realized 2D TIs, also called QSHE systems, where the helical edge states propagate at the edges of the sample, were theoretically proposed [4,20] and experimentally observed in HgTe/CdTe [5] and InAs/GaSb/AlSb quantum well structures [21]. 2D TIs possess one-dimensional (1D) metallic edge states but insulating 2D surface states. The 2D TI can be further generalized to 3D TI in theory [22], and in 2008 the first 3D TI was experimentally discovered in $Bi_{1-x}Sb_x$ alloys [23]. Soon after, the well-known $Bi_2Se_3$ family of thermoelectric materials was also found as 3D TI [24,25]. Up to now, the $Bi_2Se_3$ family TIs are the most popular TI materials due to their stoichiometric chemical composition, large bulk gap and single Dirac surface state. These 3D TIs possess 2D metallic Dirac surface states (SSs) but insulating bulk bands (Fig. 2). The inversion between the conduction band (CB) and valence band (VB) in a 3D TI caused by strong spin-orbit coupling results in the existence of the Dirac SS, which is further protected by the TRS at Γ point of the Brillouin zone.

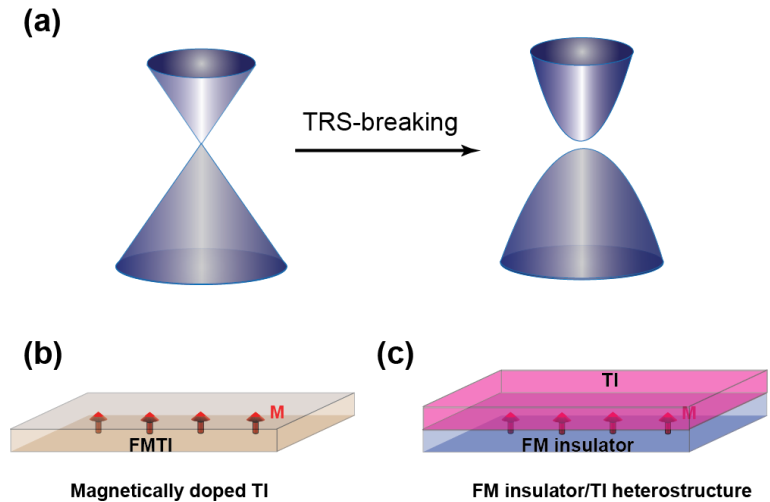

**Fig 3| Time-reversal symmetry (TRS)-breaking of TI.** (a) Surface states diagram in TI and TRS-breaking TI. (b, c) Sample structures of (b) magnetically doped TI films and (c) a FM insulator / TI heterostructure. Adapted from Ref. [26].

Breaking TRS of a TI through either magnetic doping or magnetic adlayer will open up a mini-gap of the Dirac SS (Fig. 3a) and lead to many exotic quantum phenomena, such as QAHE [19,27], topological magnetoelectric effects [27] and image magnetic monopole [28], etc. This *Topical Review* focuses on the QAHE in TRS-breaking TI materials.

## 2. Theoretical prediction of QAHE based on TI

The revolutionary discovery of TI materials made the search for the QAHE practical. Soon after the realization of QSHE in HgTe/CdTe quantum well structure [5], Liu et al. predicted that QAHE can be realized by introducing Mn ions into HgTe quantum well structure [29]. When one spin channel in QSHE is suppressed due to TRS breaking by the magnetization, the other spin channel should exhibit QAHE (Fig. 1e). However, the Mn-doped HgTe cannot form FM order due to the quenched Mn moments, prohibiting the realization of QAHE in this system. In 2009, the discovery of the $Bi_2Se_3$ family TI brought forth more candidates for the realization of QAHE [24,25]. Almost at the same time, Yu et al. theoretically predicted that FM order in TI can be induced by van Vleck mechanism, which is a $2^{nd}$-order perturbation effect (Fig. 4d). In other words, the FM order in TI can be realized in an insulating state without mediating through free carriers as the Ruderman-Kittel-Kasuya-Yosida (RKKY) case (Fig. 4c) [19]. They found that when the thickness of a magnetic TI film is reduced into a 2D hybridization regime, the QAHE can be realized as long as the hybridization gap is smaller than the FM exchange energy. In this section, we will first list the theoretical predictions in TRS-breaking TIs and then introduce the pathway to break the TRS of TIs.

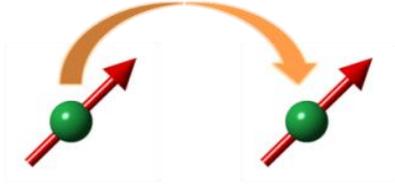
**(a) Ferromagnetic Insulator:**

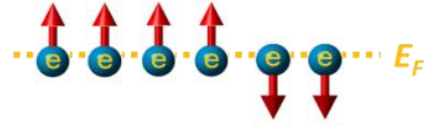
**(b) Ferromagnetic Metal:**

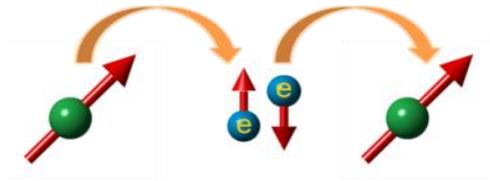
**(c) RKKY Magnetism :**

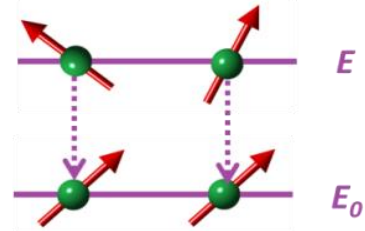
**(d) van Vleck Magnetism:**

**Fig. 4 |Common types of magnetism.** (a) FM insulator, which is direct Heisenberg exchange coupling between magnetic ions. (b) FM metal, which relies on different occupations between spin-up and spin-down electrons. (c) RKKY interaction in diluted magnetic semiconductors (DMSs), where magnetic ions are coupled indirectly through itinerant electrons. (d) van Vleck magnetism, which is a 2$^{nd}$ order perturbation effect.

## 2.1 QAHE in magnetic two-dimensional TIs

The realization of the QSHE in a HgTe/CdTe quantum well structure [5] brought a new platform to realize the QAHE, since the QSHE can be considered as two copies of the QAHEs with the opposite quanta of the Hall conductance. In 2008, based on the 2D TI four-band model that was first developed by Bernevig et al. [4], Liu et al. predicted that the QAHE can be realized by introducing Mn ions into a HgTe quantum well structure [29]. Due to the magnetization, a phenomenological term describing the spin splitting induced by intrinsic magnetization should be introduced. The phenomenological spin splitting term is:

$$H_s = \begin{pmatrix} G_E & 0 & 0 & 0 \\ 0 & G_H & 0 & 0 \\ 0 & 0 & -G_E & 0 \\ 0 & 0 & 0 & -G_H \end{pmatrix},$$

(1)

where spin splitting is $2G_E$ for the spin $\pm 1/2$ $\left| E1, \pm \right\rangle$ bands and $2G_H$ for the spin $\pm 3/2$ $\left| H1, \pm \right\rangle$ bands (Fig. 5). The energy gap of the up/down spin block is given by $E_g^{\pm} = 2M_0 \pm \left( G_E - G_H \right)$, respectively, where $2M_0$ is the gap size at the $\Gamma-$point prior to spin splitting. In order to obtain the QAHE, two conditions should be satisfied, $E_g^+ E_g^- = \left( 2M_0 + G_E - G_H \right)\left( 2M_0 - G_E + G_H \right) < 0$ and $\left( 2M_0 + G_E + G_H \right)\left( 2M_0 - G_E - G_H \right) > 0$. Combining these two conditions, if $G_E G_H < 0$, i.e. the spin splitting for $\left| E1, \pm \right\rangle$ and $\left| H1, \pm \right\rangle$ have the opposite spins (Fig 5a), the spin-down states $\left| E1, - \right\rangle$ and $\left| H1, - \right\rangle$ touch each other and then enter into the normal regime, and only the spin-up states change from a QSH state to a QAH state. It can also be understood using the edge state picture as shown in Fig. 5b. If $G_E G_H > 0$, two different spin blocks $\left| E1, + \right\rangle$ and $\left| H1, - \right\rangle$ will cross each other without opening a gap (Fig. 5c), and hence cannot create a QAH state. In other words, the condition $G_E G_H < 0$ is a prerequisite for QAHE. As mentioned earlier, a Mn-doped HgTe/CdTe quantum well was predicted to be a QAH insulator. However, due to the inaccessibility of FM order, an external magnetic field had to be applied, hence obstructing the confirmation of the true QAHE in Mn-doped HgTe system.

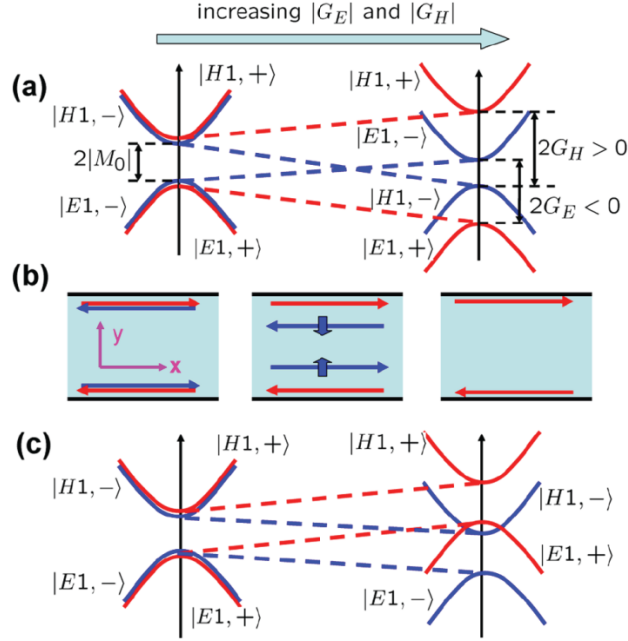

**Fig. 5| Evolution of band structure and edge states upon increasing the spin splitting.** For (a) $G_E < 0$ and $G_H > 0$, the spin-down states $|E1, -\rangle$ and $|H1, -\rangle$ in the same block of the Hamiltonian (1) touch each other and then enter the normal regime. But for (c) $G_E > 0$ and $G_H > 0$, gap closing occurs between $|E1, +\rangle$ and $|H1, -\rangle$ in the bands, which belong to different blocks of the Hamiltonian, and thus will cross each other without opening a gap. In (b) we show the behavior of the edge states during the level crossing in the case of (a). Adapted from Ref. [29].

Alternatively, another 2D TI, InAs/GaSb quantum wells with magnetic doping was also predicted to be a QAH insulator [30]. There are many advantages in this system compared to Mn-doped HgTe. First and foremost, the FM Curie temperature ($T_C$) in this system can be significantly enhanced due to the nontrivial band structures [30]. Secondly, high-quality (In,Mn)As/GaSb heterostructure have been successfully fabricated by molecular beam epitaxy (MBE) [31]. Up to now, there are no relevant experimental results on the QAHE in magnetically-doped InAs/GaSb quantum wells, probably due to the limitation of sample quality, where due to the weaker spin-orbit coupling in III-V semiconductors and resulting small bandgap, the dopants may have indirect influences to destroy QAHE.

Besides the above QAHE starting from 2D TI including HgTe quantum well [5] and InAs/GaSb quantum wells [20], from the theoretical point of view, we can even realize the QAHE in other 2D materials as long as the spin-splitting is sufficiently strong. In these 2D materials, the strong spin-splitting will cause the band inversion with single (up or down) spin state. To date, there are several proposals for the realization of the QAHE in graphene [32-37], silience [38,39], Bi (111) bilayers [40,41] and other honeycomb lattice [8]. However, there are no any relevant experimental results on the QAHE of those 2D materials, probably due to the difficulty to realize magnetic doping and the weak proximity coupling with the ferromagnetic insulators.

## 2.2 QAHE in TI/ferromagnetic insulator heterostructures

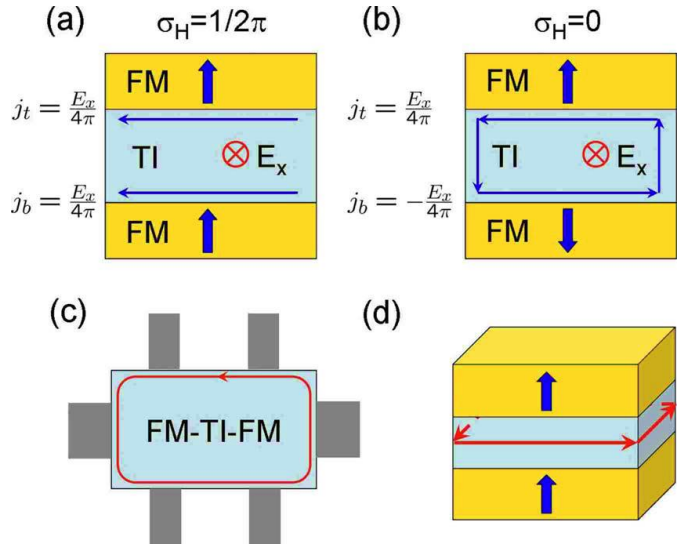

**Fig. 6| The QAHE in FM insulator/TI/FM insulator sandwich heterostructure.** (a) and (b) the electric field $E_x$ with direction into the paper and the induced Hall currents $j_t$ and $j_b$ for parallel and antiparallel magnetizations, respectively. In (b) the Hall currents on the two surfaces are opposite and form a circulating current. (c) A top-down view of the device for Hall measurements. The gray regions are leads that contact to the surface of the topological insulator. (d) For the case of parallel magnetization, the chiral edge states are trapped on the side surfaces of the TI. These carry the quantized Hall current. Adapted from Ref. [27].

In a FM insulator/TI/FM insulator sandwiched heterostructure (Fig. 6), the QAHE can also be realized. The parallel magnetization between the two FM insulator layers can lead to a quantized Hall conductance, which vanishes when the magnetization becomes antiparallel [27]. The QAHE here has the same chiral edge states as the QHE (Fig. 6d). However, all the known FM insulators (e.g. YIG, EuS, GdN, and EuO) have the in-plane easy magnetization axis, hindering the observation of QAHE in the sandwiched heterostructure. Although X. Liu et al. pointed out that the QAHE can also still be induced by in-plane magnetization [42], up to now, no relevant experiments have been reported.

## 2.3 QAHE in magnetic three-dimensional TIs

Subsequently, the tetradymite $Bi_2Se_3$ family of thermoelectric materials was found to be 3D TI with the bulk gap as large as 0.3 eV [24,25]. Moreover, well-controlled layer-by-layer MBE thin film growth of 3D TI was achieved [43-45]. Yu et al. predicted that thin films of 3D TI doped with proper TM elements (Cr or Fe) support the QAH state [19]. First, the FM order in transition metal doped 3D TI is mediated by van Vleck mechanism, which is different from the RKKY FM in the Mn-doped GaAs system [46,47]. The van Vleck FM order in TM doped 3D TI offers the possibility of an *insulating* FM order, which is exactly the requirement of QAHE.

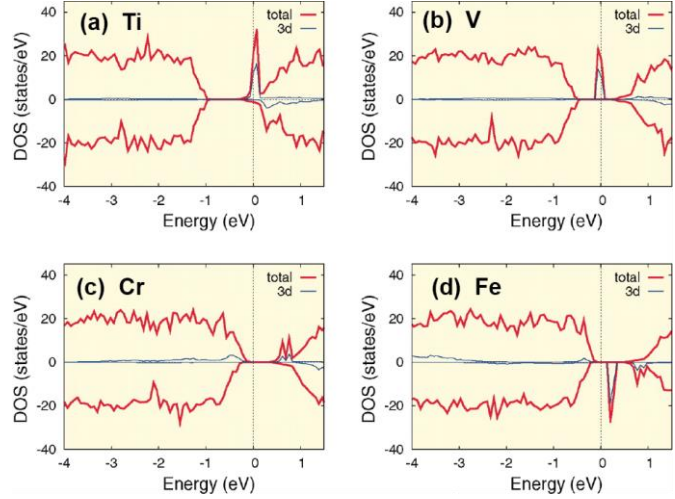

**Fig. 7| The calculated density of states (DOS) for Bi₂Se₃ doped with different transition metal elements.** The Fermi level is located at energy zero, and the positive and negative values of DOS are used for up and down spin, respectively. The blue lines are projected partial DOS of the *3d* states of transition metal ions. This predicts that Cr (a) or Fe doping (c) should give rise to the insulating magnetic state. Adapted from Ref. [19].

First-principles calculations suggest that an insulating magnetic state should be obtained for Fe or Cr doping, whereas the states are metallic for Ti or V doping cases, as shown in Fig. 7 [19]. Unlike the QAHE in Mn-doped HgTe system [29], in Fe or Cr doped 3D TI thin films, no matter whether two subbands with opposite parity are inversed or not, a strong enough exchange field will induce QAHE thanks to the presence of the $\sigma_z$ matrix in the exchange field ($gM\sigma_z$) and the opposite signs of the Zeeman coupling terms in the upper and lower blocks of the effective Hamiltonian [19].

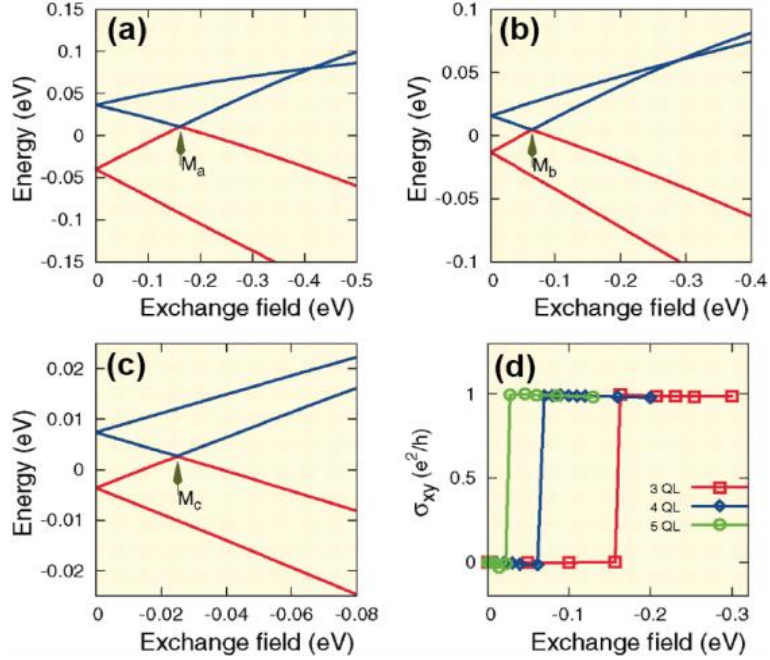

**Fig. 8| The QAH conductance in magnetic 3D TI.** (a to c) The lowest subbands at the Γ point plotted versus the exchange field, for Bi$_2$Se$_3$ films with thicknesses of three, four, and five quintuple layers (QL), respectively. The red lines denote the top occupied states; blue lines, bottom unoccupied states. With increasing exchange field, a level crossing occurs (arrows), indicating a quantum phase transition to QAH state. (d) The calculated Hall conductance for three, four, and five QL Bi$_2$Se$_3$ films versus the ferromagnetic exchange field. The Hall conductance is zero before the QAH transition, but σ$_{xy}$ = $e^2/h$ afterward. Adapted from Ref. [19].

Yu et al. plotted the lowest four subbands levels at the Γ−point as a function of the exchange field (Figs. 8a - 8c) [19]. The level crossings between the lowest conduction bands and valence bands are found for 3 quintuple layer (QL), 4QL and 5QL Bi$_2$Se$_3$ thin films, indicating the QAH state in these three doped-Bi$_2$Se$_3$ films (Fig. 8d). According to this proposal, from an experimental point of view, to realize QAHE in thin magnetically-doped thin films, three conditions should be satisfied simultaneously:

(a) Uniform-thickness thin films should be grown on insulating substrate for valid transport measurement.

(b) The long-range FM order should be introduced in thin 3D TI films with transition metal doping, and this FM order must be able to hold even in the insulating state; according to Yu's proposal, the Cr and Fe are possibly two of the best dopant options.

(c)The chemical potential of the sample must be finely tuned into the gap induced by FM. In the next section, we will introduce our route to the experimental observation of QAHE.

## 2.4 The pathways to break the time-reversal symmetry in TIs

Breaking TRS is the prerequisite for realizing the QAHE in TI materials [19,27]. The spontaneously broken TRS states can be experimentally introduced into a material by FM ordering. This may usually be achieved by two methods that are described in the following paragraphs: (1) conventionally by doping with certain magnetic ions [19]; and (2) by FM proximity exchange coupling effect [27]. In both cases, it is expected that an exchange gap opens in the Dirac SS (Fig. 3a). However, up to today, the QAHE has only been observed by (1), probably thanks to a stronger FM order and larger gap opening.

Similar to conventional diluted magnetic semiconductors (DMSs) [46,47], impurity doping using transition metal (TM) elements (e.g., Cr, V, Mn) is a convenient approach to induce long-range FM order in TIs [48-51]. Many recent experiments, from angle-resolved photoemission spectroscopy (ARPES) to electrical transport measurements, have been devoted to the study of magnetically-doped TIs of the $Bi_2Se_3$ family, including $Bi_2Se_3$, $Bi_2Te_3$, and $Sb_2Te_3$ [6,7,51-58]. In magnetically-doped $Bi_2Te_3$ and $Sb_2Te_3$, both transport and magnetometry measurements have shown clear long-range FM order in several cases [6,7,55,58], which also resulted in the observation of QAHE [6,7]. The Zeeman gap opening at the Dirac point, however, has not been resolved by ARPES or scanning tunneling spectroscopy (STS), possibly due to their low $T_C$ and small bandgap. On the other hand, for magnetically-doped $Bi_2Se_3$, the SS Zeeman gap has been reported in several ARPES studies [52,53,56], with gap size varying from several tens to a hundred meV, implying strong FM ordering. However, transport and magnetometry measurements are only able to show very small or no magnetic hysteresis under perpendicular magnetic field [52-54,56], hindering further progress toward observing QAHE in Se-based TIs.

Proximity-induced FM in a TI is attainable in a heterostructure of TI and a ferromagnet obtained by thin-film deposition technique such as MBE [27]. Commonly used FM layers, such as Fe, Co, and Ni are all metallic. However, they do not grow with high quality on top of a TI due to the low surface energy and strong oxidizing nature of the TI's Te and Se elements with metals. Most importantly, the conduction will be dominated by the metallic FM layers. Magnetic adatoms, such as Fe or Mn, may be used to avoid forming a continuous conducting layer, whereas these adatoms act as magnetic impurities which lead to spin scattering and become detrimental to TI SSs. Therefore, inert FM insulators, such as EuS and YIG [59-61], are highly advantageous. The short-range nature of magnetic proximity coupling allows the FM interaction to be restricted to TI SSs only. In this case, the all-important advantage is that the TRS breaking happens mostly at the interface between the TI and the FM insulators with neither altering bulk states nor introducing defects. However, in general, thin films of FM insulators show in-plane magnetization, hindering the observation of the QAHE.

## 3. The route to the experimental observation of QAHE

### 3.1 Quantum well TI films on insulating substrate

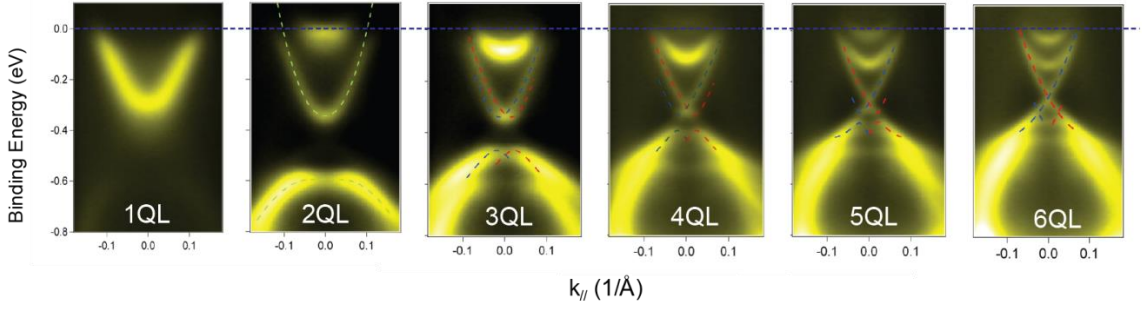

**Fig. 9 |ARPES spectra of 1QL to 6QL Bi$_2$Se$_3$ films on graphene terminated 6H-SiC (0001) substrates along $\overline{\Gamma} - \overline{K}$ direction.** All the measurements were taken at room temperature. Adapted from Ref. [43].

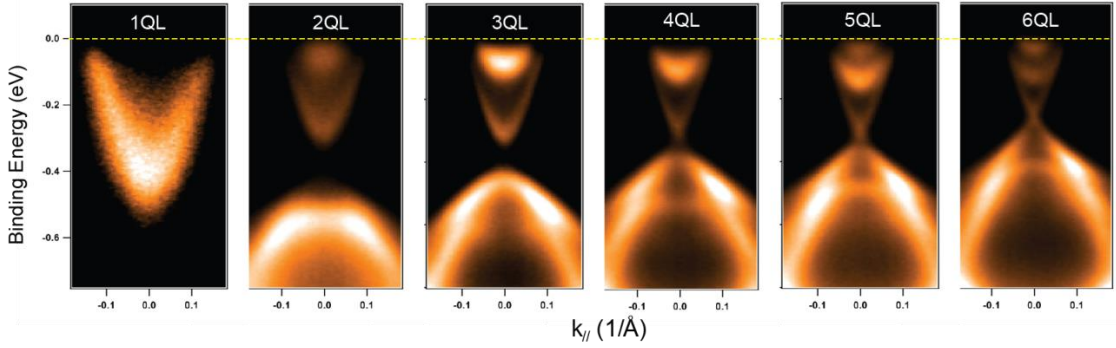

**Fig. 10| ARPES spectra of 1QL to 6QL Bi$_2$Se$_3$ films on sapphire (0001) substrates along $\overline{\Gamma} - \overline{K}$ direction.** All the measurements were taken at room temperature. Adapted from Ref. [62].

Starting with the growth of quantum well films of Bi$_2$Se$_3$, we successfully grew uniform and well-defined thickness films on bilayer graphene terminated 6H-SiC (0001) substrates (Fig. 9) [43]. The ARPES band maps clearly reveal a gap opening at the Dirac surface band in 5QL or thinner films due to the hybridization of top and bottom surface states. Later, in order to obtain valid intrinsic transport contributions from TI and exclude the substrate effects, we tested various substrates including sapphire (0001), SrTiO$_3$ (111), mica and insulating Si (111) etc., and found that the TI films on sapphire (0001) and SrTiO$_3$ (111) substrates have almost the same quality as the films on the bilayer graphene substrates. The $Bi_2Se_3$ thin film's thickness-dependent ARPES band maps on sapphire (0001) show similar surface band structure with the TI films on graphene (Fig. 10) [62]. In the spectra for 1 QL film, only one single parabolic electron band can be observed. At 2 QL, the spectra show a hole band and an electron band separated by a gap of ~245 meV. With increased thickness, the hole band and the electron band gradually approach to each other, and finally merge into a Dirac cone at 6 QL. The only difference is that the Rashaba-type splitting from 3 QL to 5 QL cannot be resolved. The Rashba splitting results from the different chemical potentials between the interface and surface of the film induced by the substrate. Its absence suggests that the $Bi_2Se_3$ films grown on sapphire (0001) have less asymmetry, which might be attributed to the rather large energy gap of the $Al_2O_3$ substrate. For $Sb_2Te_3$, according to STS studies, gapless Dirac SSs have already been developed in a 4 QL film, which shows a zero-mode ($n$=0) Landau level in a magnetic field, whereas a 3 QL film shows a SS gap of ~50 meV [63]. For $Bi_2Te_3$, 3D-2D crossover occurs between 1 QL and 2 QL [44]. It is clear that quantum well growth of $Bi_2Se_3$, $Sb_2Te_3$, and $Bi_2Te_3$ on insulating substrates satisfies the first requirement for realizing QAHE [19].

### 3.2 Inducing FM order into the quantum well TI films

According to the prediction [19], next we will introduce features of the FM state in the TI via conventional TM elements doping. Among the three members of the $Bi_2Se_3$ family TIs (including $Bi_2Se_3$, $Bi_2Te_3$ and $Sb_2Te_3$), $Bi_2Se_3$ is naturally the first choice due to its largest bulk gap (~0.3 eV) and the location of Dirac point within the bulk gap. However, no long-range FM order has been observed in Cr- and Fe-doped $Bi_2Se_3$ down to $T$=1.5 K [53,56]. In Cr-doped $Bi_2Se_3$ thin films, we found a crossover between weak antilocalization and weak localization [57,64] and chemical potential dependent gap-opening by aggregated substitutional Cr ions [56]. The short-range FM order formed by aggregated substitutional magnetic elements can explain why the gapped Dirac SS is observed in the absence of long-range FM order in Cr- and Fe-doped $Bi_2Se_3$. The absence of a long-range FM state due to inhomogeneity rules out the Cr- and Fe-doped $Bi_2Se_3$ as QAHE candidates.

Compared with Cr-doped $Bi_2Se_3$, the Cr-doped $Bi_2Te_3$ exhibits a very different transport behavior. At $T$=1.5 K, the Hall traces of Cr-doped $Bi_2Te_3$ (Fig. 11a, $x$=0) have a pronounced hysteresis loop, indicating a good long-range FM order [58]. To understand the contrasting magnetotransport properties of Cr-doped $Bi_2Se_3$ and $Bi_2Te_3$, Zhang et al. studied 8 QL Cr-doped $Bi_2(Se_xTe_{1-x})_3$, an isostructural, isovalent mixture of $Bi_2Se_3$ and $Bi_2Te_3$. The Hall traces of all the films measured at $T$ = 1.5 K are summarized in Fig. 11a, revealing a highly systematic evolution of coercivity ($H_c$) and the AHE resistivity $\rho_{yx}^0$ at zero applied magnetic field. They further constructed a magnetic phase diagram by plotting the $\rho_{yx}^0$ values as a function of the Se concentration $x$. At the base temperature $T$ = 1.5 K, the phase diagram is separated into two distinct regimes: a FM phase with $\rho_{yx}^0$>0 and a paramagnetic phase with $\rho_{yx}^0$<0 (Fig. 11b). Since the phase transition is driven by the change of chemical composition in the ground state instead of thermal fluctuations, it is a quantum phase transition as confirmed by ARPES measurements and density functional theory calculations [58].

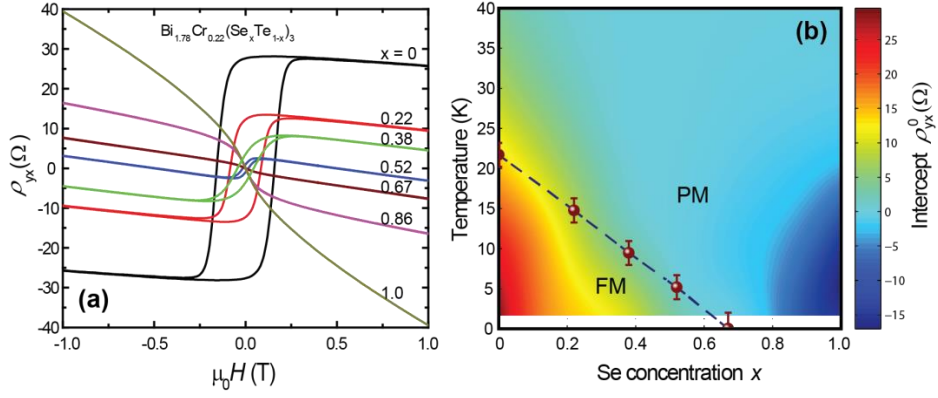

**Fig. 11| Topological phase transition in Cr-doped Bi$_2$(Se$_x$Te$_{1-x}$)$_3$.** (a) Systematic evolution of the Hall effect for all the samples ($0 \leq x \leq 1$) in 8 QL Bi$_{1.78}$Cr$_{0.22}$(Se$_x$Te$_{1-x}$)$_3$ measured at $T$ = 1.5 K. (b) Magnetic phase diagram of Bi$_{1.78}$Cr$_{0.22}$(Se$_x$Te$_{1-x}$)$_3$ summarizing the intercept $\rho^0_{yx}$ as a function of $x$ and $T$. The $T_C$ of the FM phase is indicated by the solid symbols. Adapted from Ref. [58].

Before the conceptual realization of TIs, magnetically doped Sb$_2$Te$_3$ and Bi$_2$Te$_3$ have been studied extensively as important thermoelectric materials. Long-range FM order in these materials can be easily realized via Cr and V doping [48-50]. First-principle calculations originally predicted there is no QAH state in V-doped TI system due to the existence of *d* electron impurity band at Fermi level [19]. In addition, compared to Cr-doped Bi$_2$Te$_3$ (Fig. 11a), Cr-doped Sb$_2$Te$_3$ has much better FM order as seen by a sharper magnetization reversal (Fig. 12), inspiring us to choose Cr-doped Sb$_2$Te$_3$ as the parent material for realizing the QAHE at the beginning, as discussed in Section 4. Quite strikingly, 2 years after the first observation QAHE in Cr-doped TI [6], we discovered a much better QAHE in V-doped TI system [7]. We will postpone the QAHE in V-doped TI in detail in Section 5.

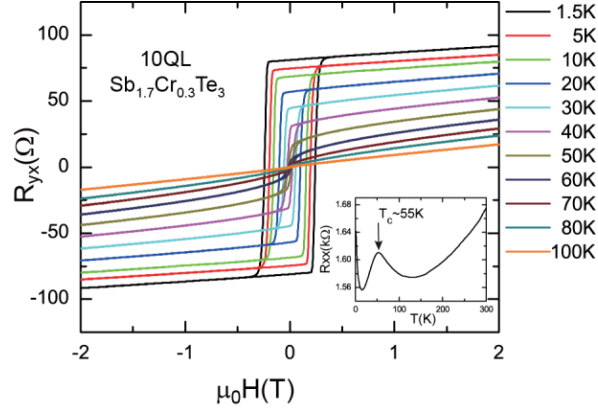

**Fig. 12| The FM response of Cr-doped Sb$_2$Te$_3$ thin films.** Magnetic field (μ$_0$H) dependent Hall resistance ($R_{yx}$) of the 10 QL Sb$_{1.7}$Cr$_{0.3}$Te$_3$ films at various temperatures. Inset, the longitudinal resistance ($R_{xx}$) vs $T$ curve. Adapted from Ref. [65].

### 3.3 Tuning the chemical potential in the surface states exchange gap

The final step of realizing QAHE is to tune the chemical potential exactly into the exchange gap induced by FM order. The first method we used was chemical doping [66]. Bi$_2$Te$_3$ and Sb$_2$Te$_3$ belong to the same space group and share very close lattice constants [24]. Beyond that, Bi$_2$Te$_3$ is intrinsically n-type owing to the presence of Te vacancies [44], while Sb$_2$Te$_3$ is generally p-type due to antisite defects [45]. Inspired by the idea of energy band engineering in conventional semiconducting (Al,Ga)As [67], by alloying Bi$_2$Te$_3$ and Sb$_2$Te$_3$ into (Bi,Sb)$_2$Te$_3$ with a certain Bi and Sb ratio, one can tune the chemical potential crossing the Dirac point. Both ARPES results and magnetotransport measurements have confirmed the effectiveness of Dirac band engineering in (Bi, Sb)$_2$Te$_3$ compounds [66].

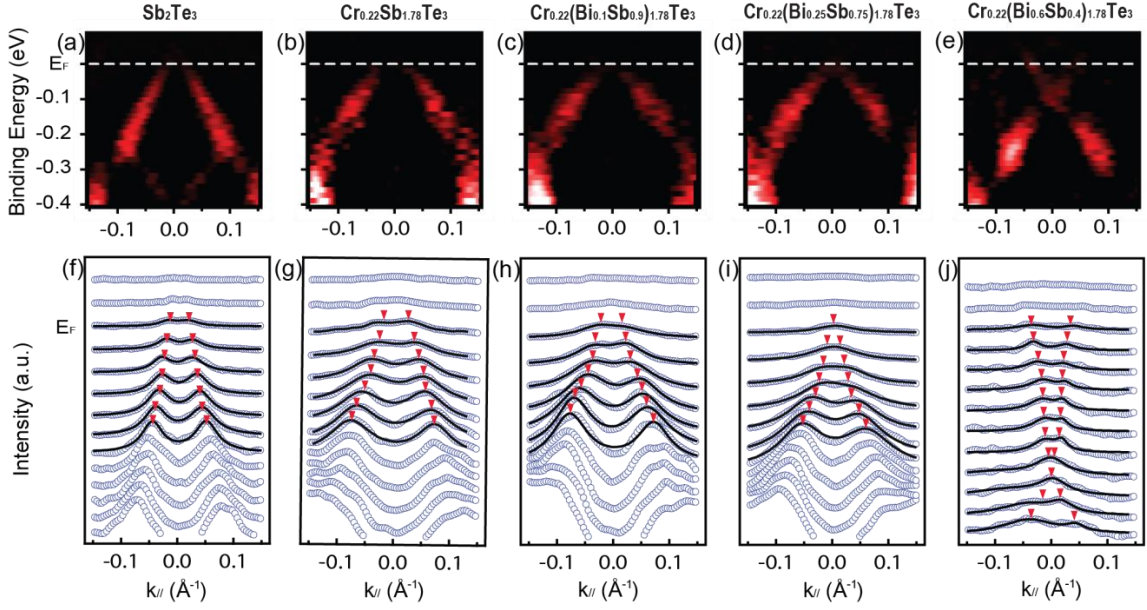

**Fig. 13| Influence of Cr and Bi dopants on the Dirac surface states of Sb$_2$Te$_3$.** ARPES grey-scale bandmaps of a Sb$_2$Te$_3$ film (a) and the Cr$_{0.22}$(Bi$_x$Sb$_{1-x}$)$_{1.78}$Te$_3$ films with $x = 0$ (b), $x = 0.1$ (c), $x = 0.25$ (d), and $x = 0.6$ (e). (f)-(j) ARPES momentum distributed curves (MDCs) corresponding to (a)-(e). In (f)-(j), the blue empty circles are data points; the black solid lines are results of fitting with a double Lorentz peak; the red triangles indicate the fitted peak positions. Adapted from Ref. [55].

The Dirac band engineering can also be realized in magnetically-doped TI. By substituting Sb with Bi in Cr-doped Sb$_2$Te$_3$, the chemical potential can be tuned across the Dirac point. Fig. 13 displays the ARPES bandmaps of a pure Sb$_2$Te$_3$ film and four Cr$_{0.22}$(Bi$_x$Sb$_{1-x}$)$_{1.78}$Te$_3$ films with systematically varied Bi:Sb ratios. The two linearly dispersed bands crossing the Fermi level shown in the spectra for Sb$_2$Te$_3$ are Dirac surface bands. The Dirac point is located above the Fermi level, indicating $p$-doping of the sample. Even at a high Cr magnetic doping level (>10%), the Dirac surface bands can still be clearly resolved. Doping with Bi has little influence on the band dispersion of the Dirac surface states, as shown in the spectra for all Bi concentrations. On the other hand, the Fermi level is shifted upwards with increasing Bi content, from below the Dirac point to near the charge neutral point around $x = 0.25$, and then to above the Dirac point. The existence of Dirac SSs in all the samples suggests that the materials remain 3D TIs even at such a high doping level, without showing any topological phase transition like that in Cr-doped $Bi_2Se_3$ [58]. This property is crucial for the observation of QAHE.

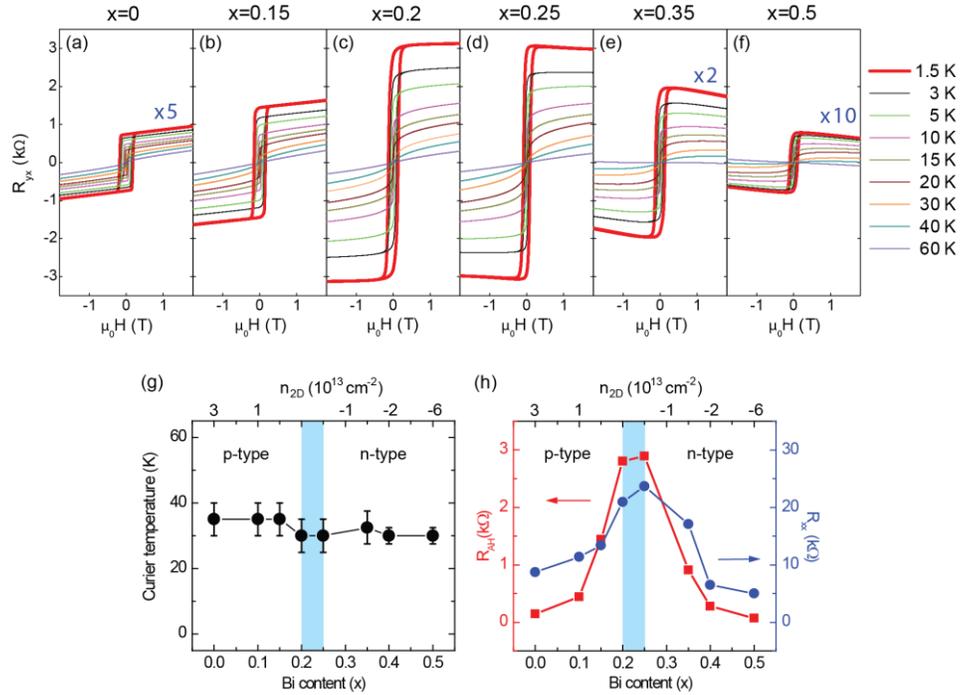

**Fig. 14| Magnetic properties of Cr-doped $(Bi_xSb_{1-x})_2Te_3$ films at different carrier density and type via chemical doping.** (a)-(f) Magnetic field dependent Hall resistance $R_{yx}$ of $Cr_{0.22}(Bi_xSb_{1-x})_{1.78}Te_3$ films with $x = 0$ (a), $x = 0.15$ (b), $x = 0.2$ (c), $x = 0.25$ (d), $x = 0.35$ (e), $x = 0.5$ (f) at different temperatures. (g) Dependence of Curie temperature ($T_C$) on Bi content ($x$) (bottom axis) and carrier density (top axis). Error bars represent the temperature spacing in $R_{AH}$-$T$ measurements. (h) Dependence of anomalous Hall resistance ($R_{AH}$) (red solid squares) and longitudinal resistance ($R_{xx}$) (blue solid circles) at 1.5 K on Bi content ($x$) (bottom axis) and carrier density (top axis). Adapted from Ref. [55].

Fig. 14 shows the magnetic field dependent Hall resistance of 5QL $Cr_{0.22}(Bi_xSb_{1-x})_{1.78}Te_3$ films grown on sapphire (0001) with different Bi:Sb ratios and constant Cr-doping level. At $T = 1.5$ K, all films show nearly square-shaped AHE hysteresis loops, suggesting strong ferromagnetism with perpendicular axis magnetization. With increased Bi concentration, the ordinary Hall effect evolves from positive to negative, showing carrier type changing from *p*- to *n*-type, without impacting the FM order of Cr-doped $(Bi_xSb_{1-x})_2Te_3$. The $T_C$ remains unchanged even in the highly-insulating samples around the *p*- to *n*-transition (Fig. 14g). This clearly demonstrates a carrier-independent ferromagnetism, supporting the existence of a FM insulator phase that is likely due to a van Vleck mechanism [19]. The AH resistance, on the other hand, exhibits a dramatic and systematic change with Bi-doping. The largest AH resistance was ~3 kΩ, when the sample had the lowest carrier density (Fig. 14h). Although this number is larger than the AH resistance observed in most FM metals, it still is far from the quantized value (~25.8 kΩ). Solely by coarse adjustment of the Bi and Sb ratio during MBE sample growth, it is quite challenging to fine tune the chemical potential precisely into the surface states exchange gap. Hence, we introduced the bottom electrical field gating effect into our studies. A $Cr_{0.22}(Bi_{0.2}Sb_{0.8})_{1.78}Te_3$ film is grown on a $SrTiO_3$ (111) substrate. The high dielectric constant ε of $SrTiO_3$ at low temperature (ε~30000) allows for the possibility of applying a sufficiently large electric field to fine tune the carrier density [68]. Fig. 15a shows the hysteresis loops of $R_{yx}$ measured at 250 mK with different $V_g$. $R_{AH}$ changes dramatically with $V_g$, from 660 Ω at -210 V to 6.1 kΩ (the maximum) at 35 V, as summarized in Fig. 15b. $R_{xx}$ exhibits a peak of ~30 kΩ at 80 V. The different peak positions can be attributed to the ferroelectricity of the substrate, causing hysteresis in gate tuning [69]. The AH angle corresponding to $R_{AH} = 6.1$ kΩ is ~0.2 (the red star in Fig. 15c), indicating an unusually large AHE. The coercivity and the shape of the hysteresis loops are nearly unchanged with $V_g$, suggesting an almost unchanged $T_C$. The carrier density estimated from ordinary Hall effect shows a crossover from *p*- to *n*-type with $V_g$, but exhibits complex behavior around the crossover region due to the large negative MR with significant hysteresis. To accurately determine the carrier density around $R_{AH} = 6.1$ kΩ, a Hall measurement with magnetic field up to $\mu_0 H$=15 T was carried out at $V_g$ = 35V. The field effect experiment further demonstrates that FM order in the system is insensitive to carrier concentration, whereas AHE is greatly enhanced with reduced carrier density. The observed carrier-insensitive FM order allows for a strong signature demonstrating the existence of the QAHE in Cr-doped $(Bi_xSb_{1-x})_2Te_3$.

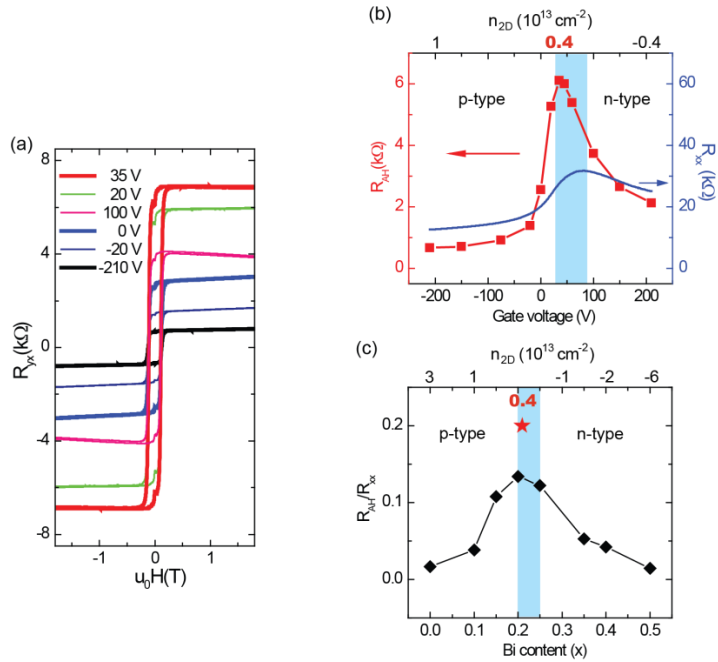

**Fig. 15| Magnetic properties of Cr-doped $(Bi_xSb_{1-x})_2Te_3$ films at different carrier density and type via bottom gating.** (a) Magnetic field dependent Hall resistance $R_{yx}$ of a 5QL $Cr_{0.22}(Bi_{0.2}Sb_{0.8})_{1.78}Te_3$ film grown on $SrTiO_3(111)$ at different back-gate voltages measured at 250 mK. (b) Dependence of $R_{AH}$ (red solid squares) and $R_{xx}$ (blue solid line) of the 5QL $Cr_{0.22}(Bi_{0.2}Sb_{0.8})_{1.78}Te_3$ film grown on $SrTiO_3(111)$ substrate on back-gate voltages at 250 mK. (c) Dependence of $R_{AH}/R_{xx}$ on Bi content ($x$) (bottom axis) and carrier density (top axis). Black solid diamonds represent the data obtained from Fig. 14. The red star represents the value obtained from a Hall measurement at 250 mK and 15 T on the $Cr_{0.22}(Bi_{0.2}Sb_{0.8})_{1.78}Te_3$ film grown on $SrTiO_3(111)$ with a back-gate voltage of 35 V. Adapted from Ref. [55].

## 4. QAHE in Cr-doped TIs

## 4.1 First experimental observation of the QAHE

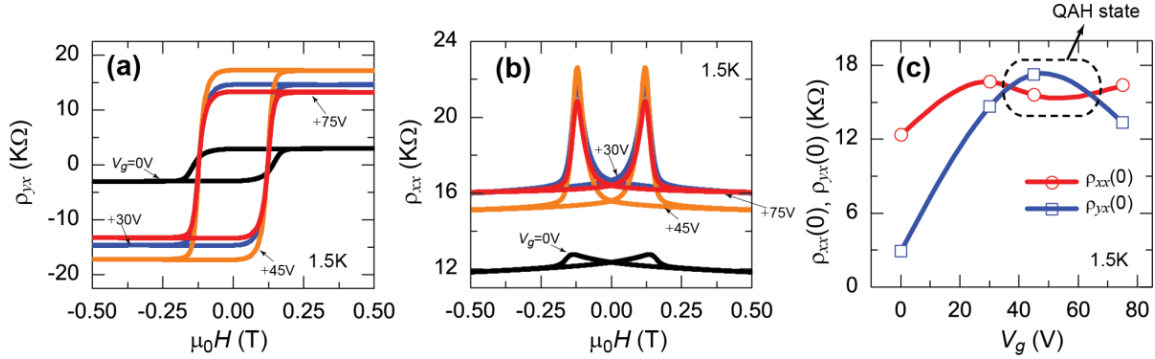

**Fig. 16| The first experimental observation of the QAH state.** (a) Magnetic field dependence of $\rho_{yx}$ at different $V_g$ values. (b) Magnetic field dependence of $\rho_{xx}$ at different $V_g$ values. (c) Dependence of $\rho_{yx}(0)$ (empty blue squares) and $\rho_{xx}(0)$ (empty red circles) on $V_g$. The purple dash-dotted oval in (c) indicates the region of $V_g$ for the charge neutral point ($V_g^0$).

As descried in Section 3, we have conducted a systematic optimization of the growth of these films. Following those growths, a better UHV environment in the MBE chamber was achieved by replacing the effusion cells and manipulators with new ones. We further realized that by slightly lowering the substrate temperature during growth, Cr ions could be implanted into the TI films with better homogeneity. The film thickness and composition were also fine-tuned and carefully checked. These efforts led to larger terraces, reduced bulk carrier density and higher mobility, which in turn resulted in larger $\rho_{yx}$ and lower $\rho_{xx}$. On October 9, 2012, we observed the QAH state for the first time in one sample with 5QL $Cr_{0.15}(Bi_{0.1}Sb_{0.9})_{1.85}Te_3$ (Fig. 16). Prior to this sample, the gate dependence of $\rho_{yx}$ always has a positive correlation with the gate dependence of $\rho_{xx}$, which cannot demonstrate the samples entering into QAH state due to the usual relationship $\rho_{yx} \propto \rho_{xx}$ [3]. However, in this sample, the behavior that $\rho_{yx}$ increased and $\rho_{xx}$ decreased near $V_g$=45V confirmed the existence of the QAH state in Cr-doped $(Bi,Sb)_2Te_3$ system. The tangent of Hall angle was about 1.1 at $T$=1.5 K.

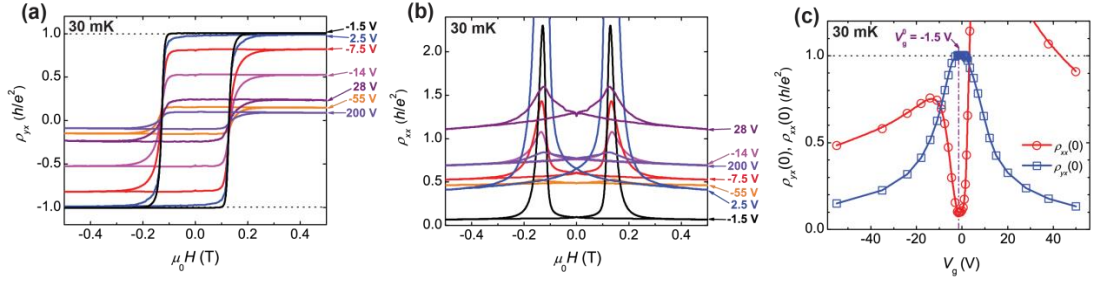

**Fig. 17| The QAHE measured at 30 mK.** (a) Magnetic field dependence of $\rho_{yx}$ at different $V_g$ values. (b) Magnetic field dependence of $\rho_{xx}$ at different $V_g$ values. (c) Dependence of $\rho_{yx}(0)$ (empty blue squares) and $\rho_{xx}(0)$ (empty red circles) on $V_g$. The vertical purple dash-dotted line in (c) indicate $V_g$ for the charge neutral point ($V_g^0$). Adapted from Ref. [6].

Ultralow temperature (~30 mK) measurements yielded exactly quantized $\rho_{yx}$ in another 5QL Cr-doped $(Bi_xSb_{1−x})_2Te_3$ thin film on $SrTiO_3(111)$ substrate. Figs.17a and 17b show the magnetic field dependence of $\rho_{yx}$ and $\rho_{xx}$, respectively, measured at $T = 30$ mK for various gate bias ($V_g$). The shape and coercivity of the hysteresis loops are nearly independent of $V_g$, indicating a carrier independent FM order. At the saturated magnetization state, $\rho_{yx}$ is nearly independent of the magnetic field, suggesting a uniform FM order and charge neutrality of the sample, whereas the AH resistance is changed dramatically with $V_g$, with a maximum value of $h/e^2$ (25.8 kΩ) occurring at $V_g$ ~1.5 V. The MR curve exhibits the typical hysteretic shape as in FM metals.

The most important observation, that the Hall resistance exhibits a distinct plateau with the quantized value $h/e^2$ at zero applied magnetic field, is shown in Fig 17c. Complementary to this $\rho_{yx}$ plateau, the longitudinal resistance $\rho_{xx}(0)$ shows a dip, reaching a value of 0.098 $h/e^2$, yielding a Hall angle of 84.4 °. For comparison with theory, we transform $\rho_{yx}(0)$ and $\rho_{xx}(0)$ into sheet conductance $\sigma_{xy}(0)$ and $\sigma_{xx}(0)$. $\sigma_{xy}(0)$ and $\sigma_{xx}(0)$ show the same behaviors on $V_g$ as the corresponding $\rho_{yx}(0)$ and $\rho_{xx}(0)$ around $V_g^0$. The $\sigma_{xy}(0)$ exhibits a notable plateau at 0.987 $e^2/h$, while the $\sigma_{xx}(0)$ exhibits a dip down to 0.096 $e^2/h$. These results demonstrate the realization of the QAHE in magnetically doped TI films. It should be noted here, the observation of the deviation of $\sigma_{xy}(0)$ from $e^2/h$ and nonzero $\rho_{xx}(0)$ and $\sigma_{xx}(0)$ can be attributed to residual dissipative channels, which is expected to vanish completely at zero temperature or in a strong magnetic field.

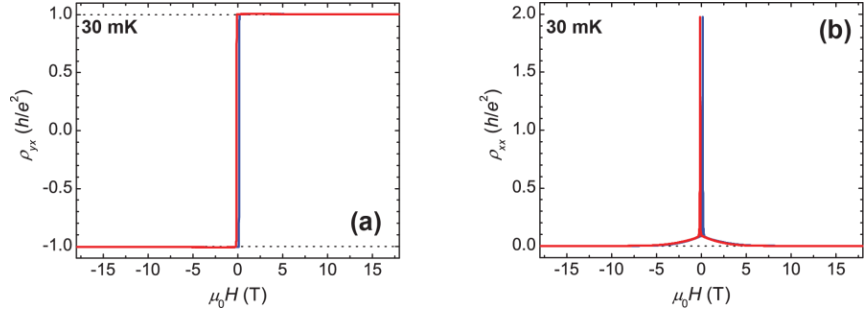

**Fig. 18| QAHE under strong magnetic field.** (a) Magnetic field dependent $\rho_{yx}$ at $V_g^0$. (b) Magnetic field dependent $\rho_{xx}$ at $V_g^0$. The measurement temperature was 30 mK. Adapted from Ref. [6].

To further confirm the QAHE, we applied a magnetic field in order to localize all possible dissipative channels in the sample. The magnetic field dependent $\rho_{yx}$ and $\rho_{xx}$ are shown in Figs. 18a and 18b. Apparently, with increasing field, $\rho_{xx}$ is further pushed toward zero. Above 10 T, $\rho_{xx}$ vanishes completely, corresponding to a perfect QH state. It is noteworthy that the increase in $\rho_{xx}$ from zero (~10 T) to 0.098 $h/e^2$ (~zero-field) is very smooth, without any trace of Shubnikov–de Haas oscillation or jumps. This suggests an absence of quantum phase transitions, and the sample remains in the same QH phase as the field sweeps from 10 T to zero field. Therefore, the complete quantization above 10 T can be only attributed to the same QAH state at zero magnetic field.

After this first observation of the QAHE in the Cr-doped (Bi,Sb)$_2$Te$_3$ system [6], several other groups also observed the QAHE in the same system [70-73]. Next, we will discuss the recent experimental progress of the QAHE in this system.

## 4.2 Trajectory of QAHE

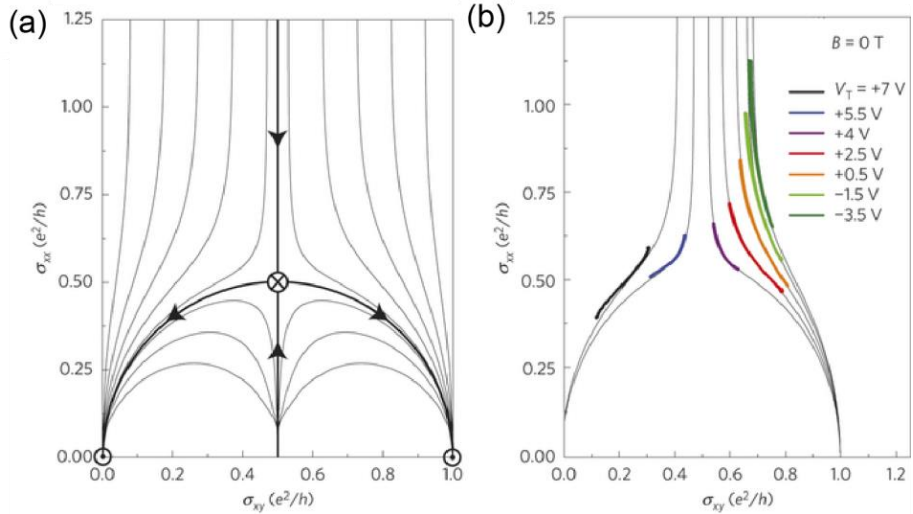

**Fig. 19| Temperature driven renormalization group flow in QAHE.** (a) Calculation of renormalization group flow for a 2D electron system with insulating and quantum Hall stable points. (b) Flow in ($\sigma_{xy}(V_T)$, $\sigma_{xx}(V_T)$) observed at B = 0 for various top gate V$_T$. On decreasing T, the system evolves towards lower $\sigma_{xx}$. The thin black lines are the calculated renormalization group flow lines from the theory in Ref. [74]. Adapted from Ref. [70].

J. G. Checkelsky et al. [70] reconfirmed the existence of the QAHE in Cr-doped (Bi,Sb)$_2$Te$_3$, and mapped out the relationships of the longitudinal and Hall conductivities as a function of top gate bias $V_t$ and temperature $T$. They showed the evidence of quantum criticality behavior while approaching the QAH state that can be described by a picture of renormalization group (RG) flow of the integer QHE [74] (Fig. 19). As $T$ is lowered, the RG flow lines of ($\sigma_{xy}(V_t)$, ($\sigma_{xx}(V_t)$) seem to converge to either one of two stable fixed points at (0,0) and ($e^2/h$,0). These flow lines in the QAHE are taken over a range of temperature from 50 mK to 700 mK, and nicely reproduce the characteristics of the calculated RG trajectories in the integer QHE at fixed chemical potential (Fig. 19a). This indicates that the QAHE and the integer QHE may have the identical critical behavior with a universal origin.

### 4.3 Nonlocal transport in QAHE

X. Kou et al. [71] fabricated 10 QL Cr-doped (Bi, Sb)$_2$Te$_3$ films on the intrinsic GaAs(111) substrates, and also observed the QAHE. Besides the conventional Hall measurements, they also made nonlocal transport measurements based on a six-terminal Hall bar in order to manifest the edge properties of the QAHE. The configuration of nonlocal transport is illustrated in Fig. 20. The first case is where the current is passed through terminals 1 and 2, while the nonlocal resistance between 5 and 4 ($R_{12,54}$) is measured. The second case is where the current is passed through terminals 2 and 6, while the nonlocal resistance between 3 and 5 ($R_{26,35}$) is measured. In the QAHE regime at the lowest temperature, it can be clearly seen that both $R_{12,54}$ and $R_{26,35}$ display square-shaped hysteresis windows, whose two-state nonlocal resistances confirm the coexistence of a QAH chiral edge channel and the additional dissipative (nonchiral) edge conduction in their sample.

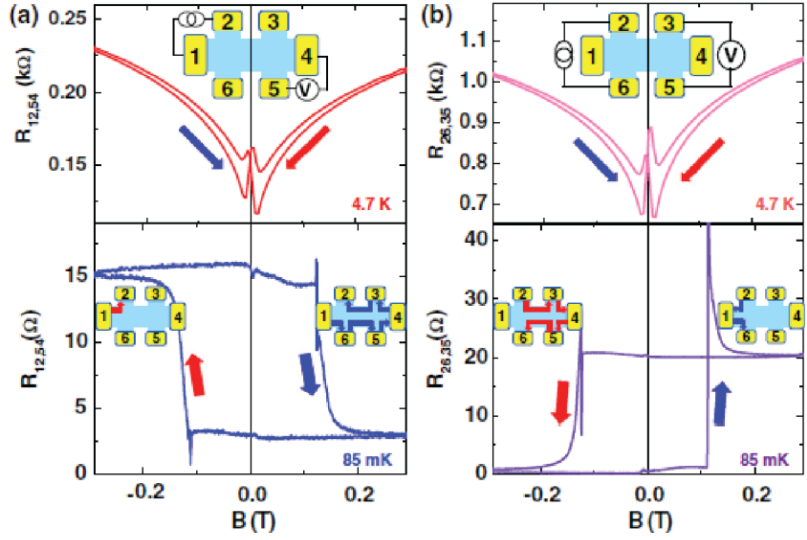

**Fig. 20| Nonlocal measurements of a six-terminal Hall bar device.** (a) Case A: The current is applied through the first to the second contact, and the nonlocal voltages are measured between the fourth and fifth contacts at $T$=85 mK and 4.7 K. (b) Case B: The current is applied through the second to the sixth contact, and the nonlocal voltages are measured between the third and fifth contacts with different magnetizations. Inset: Illustrations of the QAHE channels under different magnetizations. The red solid arrow indicates that the magnetic field is swept from positive (+z) to negative (−z), whereas the blue solid arrow corresponds to the opposite magnetic field sweeping direction. Adapted from Ref. [71].

## 4.4 Zero Hall conductivity plateau in the QAHE

If we transform the longitudinal resistance $\rho_{xx}$ and Hall resistance $\rho_{yx}$ to longitudinal conductance $\sigma_{xx}$ and Hall conductance $\sigma_{xy}$, two intermediate plateau with $\sigma_{xy}$=0 and double peaks in $\sigma_{xx}$ are clearly observed at the coercive field ($H_c$) in the QAH regime, consistent with the previous theoretical prediction. The zero Hall conductivity plateau in the QAHE can be understood as follows: In the QAH regime with the well-aligned magnetization, all the magnetic domains in the sample are also aligned along the same direction, thus there is only one pair of chiral edge channels propagating along the entire sample boundary. The Hall conductivity plateau, on the contrary, occurs around the $H_c$ during the magnetization reversal process, indicating the magnetic TI sample is driven into a large number of randomly distributed upward and downward magnetic domains. When the localization length is small enough at the lowest temperatures, the tunneling between chiral edge states at the domain wall is suppressed, and the total conduction of the QAH sample is thus minimized, so zero $\sigma_{xy}$ plateaus and the double peaks in $\sigma_{xx}$ can be observed in QAHE near $H_c$ [75]. The behavior has been independently observed by X. Kou et al. [76] and Y. Feng et al. [77].

## 5. QAHE in V-doped TIs

Based on first-principles calculations by Yu et al. [19], QAH insulators are expected to be formed in the 2D limit of a 3D TI by Cr or Fe doping, whereas metallic states are predicted to form upon Ti or V doping due to $3d$ electron impurity band formation located at the chemical potential (Fig. 7). For nearly 2 years since 2013, most QAHE experiments were dominated by Cr-doped TI thin films, since Fe-doped TIs do not exhibit any long-range FM order, while the high growth temperature of V (~2000 ℃) causes many feasibility issues. However, there is one thing that caught our attention: among the various TM atoms (Cr, V and Mn) [48-51], V-doped $Sb_2Te_3$ exhibits the most stable FM order [49], as seen by a larger hysteresis loop at fixed temperature as well as a higher $T_C$ at fixed dopant concentration (Fig. 22). This fact motivated us to explore the QAHE in V-doped TI thin films. By a similar method used for the Cr-doped QAH sample [6], a clear and high-precision QAHE was finally observed in V-doped $(Bi,Sb)_2Te_3$ epitaxial thin films [7,78]. Next we will review the more recent progress about the QAHE in V-doped TI systems.

### 5.1 Experimental verification of van Vleck type ferromagnetism

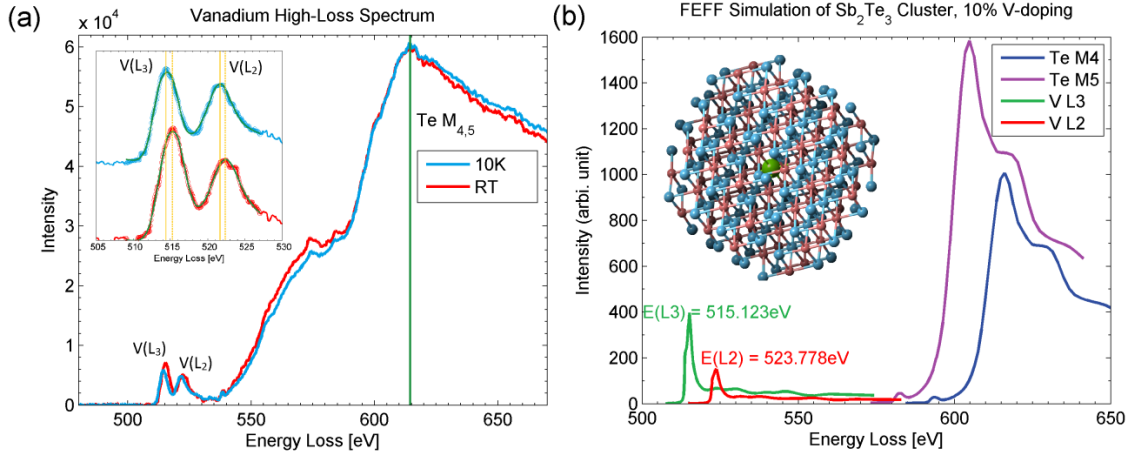

**Fig. 21 | EELS measurements of V-doped Sb$_2$Te$_3$.** (a) EELS spectra of V $L_2$, $L_3$ and Te $M_{4,5}$ edges at RT (red) and 10 K (blue) in a V-doped Sb$_2$Te$_3$ sample, normalized with the Te $M_{4,5}$ edge intensity. The energy position of Te $M_{4,5}$ edge is invariant with temperature (green line), while there is a clear redshift of vanadium's $L_3$ and $L_2$ positions (yellow lines) and a drop of the $L_3/L_2$ ratio. (b) Spectrum simulation results of a cluster, showing a comparable shift that indicates a 2$^{nd}$ order perturbation effect as the origin of the magnetism. Adapted from Ref. [79].

By low-temperature high-resolution electron energy loss spectroscopy (EELS), we demonstrated that the long-range FM order in V-doped Sb$_2$Te$_3$ has van Vleck nature, which is a type of magnetism induced by 2$^{nd}$-order energy perturbation without the need of itinerant carriers as media in the case of the RKKY interaction of usual DMS (Fig. 4). The core-level peaks ($L_3$ and $L_2$) of the V dopants in Sb$_2$Te$_3$ show unambiguous energy redshift when the sample is cooled down from room temperature to $T$=10 K (Fig. 21a). Such shift is shown to have an origin of 2$^{nd}$ order energy perturbation through Green's function's spectrum simulations (Fig. 21b) that takes into account perturbation effects automatically, which is indeed a feature of van Vleck magnetism. Moreover, since the RKKY interaction only involves itinerant carriers, such core-level shift without itinerant carriers involved verifies that the magnetism in V-doped TI has the van Vleck-type FM order in a direct manner. The van Vleck-type FM makes the realization of QAHE in V-doped Sb$_2$Te$_3$ possible.

## 5.2 The advantages of V-doped TI system

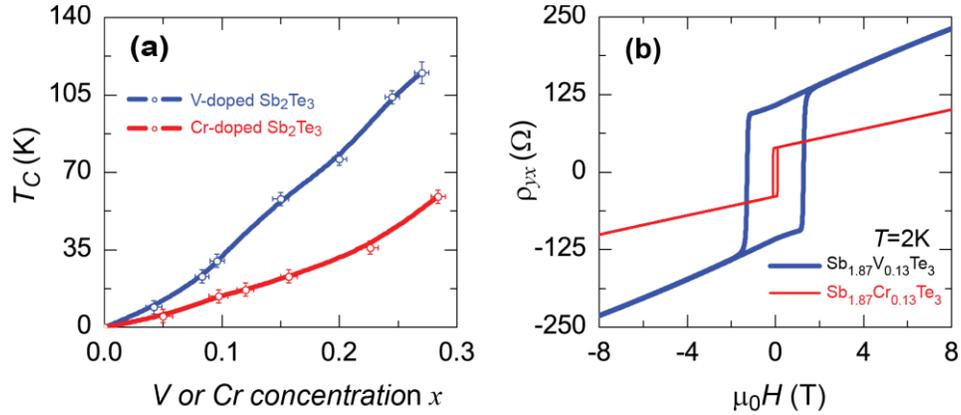

**Fig. 22| FM properties comparing Cr- and V-doped Sb₂Te₃.** (a) The Curie temperature ($T_C$) of 6QL Sb$_{2-x}$V$_x$Te$_3$ (blue circles) and Sb$_{2-x}$Cr$_x$Te$_3$ thin films (red circles). Note the factor of 2 increase in $T_C$ upon V doping. (b) The Hall traces of 6QL Sb$_{1.87}$V$_{0.13}$Te$_3$ and Sb$_{1.87}$Cr$_{0.13}$Te$_3$ thin films, measured at $T$=2 K. Note the order of magnitude increase in $H_c$ for V doping. Adapted from Ref. [7].

Before discussing the phenomena of QAHE in V-doped TI system, we would like to first remark on the advantages of V-doped Sb₂Te₃ compared with Cr-doped TI system as a candidate for realizing the QAHE. To understand the differences between these two materials, we carried out direct comparisons between the two systems. Our studies showed the following four main differences:

(a) At the same doing level $x$, the $T_C$ of V-doped Sb₂Te₃ is much larger than Cr-doped samples, almost by a factor of 2. The doping-level-dependent $T_C$ for 6QL Sb$_{2-x}$Cr$_x$Te$_3$ and Sb$_{2-x}$V$_x$Te$_3$ is shown in Fig. 22a. Since FM order is a prerequisite to reach a QAH state, the stronger FM order and higher $T_C$ of V-doping at the same doping level provides a reasonable search direction for a QAHE that is survivable at higher temperature while maintaining good sample crystallinity [80].

(b) At the sample doping level, the V-doped Sb₂Te₃ has a much greater $H_c$ than Cr-doped one. Hall traces for the doping level $x$=0.13 are shown in Fig. 22b. In both cases, the square-shaped loops indicate the long-range FM order with out-of-plane magnetic anisotropy. The $H_c$ for the Sb$_{1.87}$V$_{0.13}$Te$_3$ film is $H_c$ ~1.3 T at $T$=2 K, an order of magnitude larger than that of Sb$_{1.87}$Cr$_{0.13}$Te$_3$ with $H_c$~0.1 T. The larger $H_c$ of V-doped Sb$_2$Te$_3$ is clearly a great advantage from the perspective of a more robust FM order against external perturbations.

(c) Remarkably, at the same doping levels, the carrier density of V-doped Sb$_2$Te$_3$ is much lower than for the Cr-doped one. For instance, the $n_{2D}$ of 6QL Sb$_{1.87}$V$_{0.13}$Te$_3$ is $n_{2D}$ ~ 2.5×10$^{13}$cm$^{-2}$, only half of that for Sb$_{1.87}$Cr$_{0.13}$Te$_3$ ($n_{2D}$~4.9×10$^{13}$cm$^{-2}$). The lower $n_{2D}$ in the parent material is beneficial because it reduces the amount of Bi doping required to drive the system toward a charge neutral state [6].

(d) The Hall resistance $\rho_{yx}$ of Cr-doped Sb$_2$Te$_3$ shows nearly non-FM behavior under zero-field cooling and it must be driven into the FM state by applying an external magnetic field. In contrast, the $\rho_{yx}$ of V-doped Sb$_2$Te$_3$ shows FM behavior in the 'virgin' condition of zero-field cooling. As shown in Fig. 23a and 23c, the 'virgin' magnetized curves nearly overlap with the $\rho_{yx}$–μ$_0$$H$ hysteresis and the $\rho_{xx}$–μ$_0$$H$ butterfly structure. This is not the case for the Cr-doped Sb$_2$Te$_3$ sample. The "self-driven" magnetization in V-doped Sb$_2$Te$_3$ system enables the realization of the QAHE in an environment completely free of *any external magnetic field*.

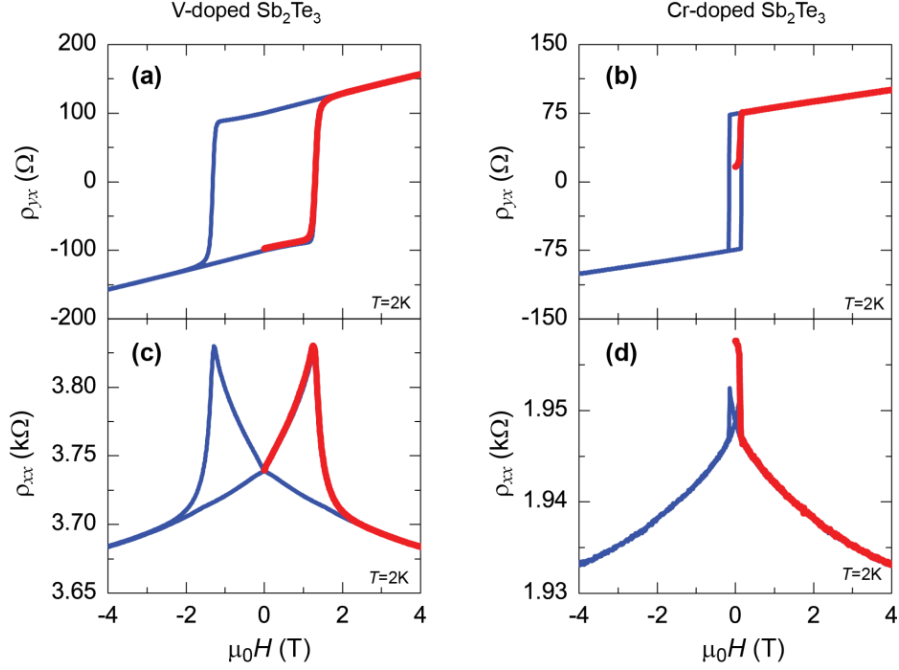

**Fig. 23| $\rho_{yx}$ and $\rho_{xx}$ hysteresis loops of V- and Cr-doped Sb$_2$Te$_3$ films.** (a, c) Field dependence of $\rho_{yx}$ and $\rho_{xx}$ of a 6QL Sb$_{1.85}$V$_{0.15}$Te$_3$ thin film. (b, d) Field dependence of $\rho_{yx}$ and $\rho_{xx}$ of a 6QL Sb$_{1.84}$Cr$_{0.16}$Te$_3$ thin film at 2 K. The red curves in (a) to (d) are the 'virgin' magnetized curves, where the samples were not exposed to external field. Adapted from Ref. [7].

An intuitive explanation accounting for all above-mentioned differences between these two materials lies in distinct magnetic moments and valence states of V and Cr ions. The saturation magnetic moment per V ion was determined to be 1.5 $\mu_B$ (Bohr magneton) in $x$=0.13 V-doped Sb$_{2-x}$V$_x$Te$_3$. This suggests that the valence state of V is a mixture of 3+ and 4+ (and possibly 5+) as it is expected to substitute for the Sb$^{3+}$ ion on the Sb sub-lattice. The extra free electrons resulting from V$^{4+}$ (or/and V$^{5+}$) tend to neutralize the $p$-type carriers, so the $n_{2D}$ of V-doped Sb$_2$Te$_3$ is expected and experimentally found to be lower than that of Cr-doped samples.

In addition, the order-of-magnitude higher $H_c$ found in the V-doped system can be explained as a much smaller magnetic moment of the V atom (~1.5 $\mu_B$) compared with the Cr atom (~3 $\mu_B$). The $H_c$ in FM films, in addition to extrinsic influences such as defects that pin domain walls, depends on the magnetic anisotropy and saturation magnetization moment ($M_s$) of the films. For magnetization reversal by the process of magnetic domain rotation, $H_c \propto \frac{K}{M_s}$, where $K$ is the total magnetic anisotropy constant, thus a higher $K$ and/or lower $M_s$ lead to a higher $H_c$. Since the demagnetization energy ($E_d$) determines the density of magnetic domain walls, and $E_d \propto M_s{}^2$, materials with a few large domains have a high $H_c$, while those with many small domains have a low $H_c$. Therefore, it is clear that V-doped $Sb_2Te_3$ with a lower $M_s$ and a higher $H_c$ have fewer magnetic domains than Cr-doped systems, indicating that the magnetic domains in V-doped $Sb_2Te_3$ are much larger in size than in Cr-doped films.

### 5.3 High-precision and self-driven QAHE

The magnetotransport on a 4 QL $(Bi_{0.29}Sb_{0.71})_{1.89}V_{0.11}Te_3$ film shown in Fig. 24 exhibits nearly ideal QAH behavior. At the charge neutral point $V_g^0$, the Hall resistance at zero magnetic field $\rho_{yx}(0)$ displays a nearly perfect quantized value of $1.00019 \pm 0.00069 h/e^2$ ($25.8178 \pm 0.0177$ k$\Omega$), while the longitudinal resistance at zero magnetic field $\rho_{xx}(0)$ is as low as $0.00013 \pm 0.00007$ $h/e^2$ (~$3.35 \pm 1.76$ $\Omega$) measured at $T$=25 mK (Figs.24b and 24c). The ratio $\rho_{yx}(0)/\rho_{xx}(0)$ corresponds to an anomalous Hall angle $\alpha$ as large as $89.993 \pm 0.004°$. The corresponding Hall conductance at zero magnetic field $\sigma_{yx}(B=0)$, is found to be $0.9998 \pm 0.0006$ $e^2/h$ and the longitudinal conductance at zero magnetic field $\sigma_{xx}(0)$ gives $0.00013 \pm 0.00007$ $e^2/h$.

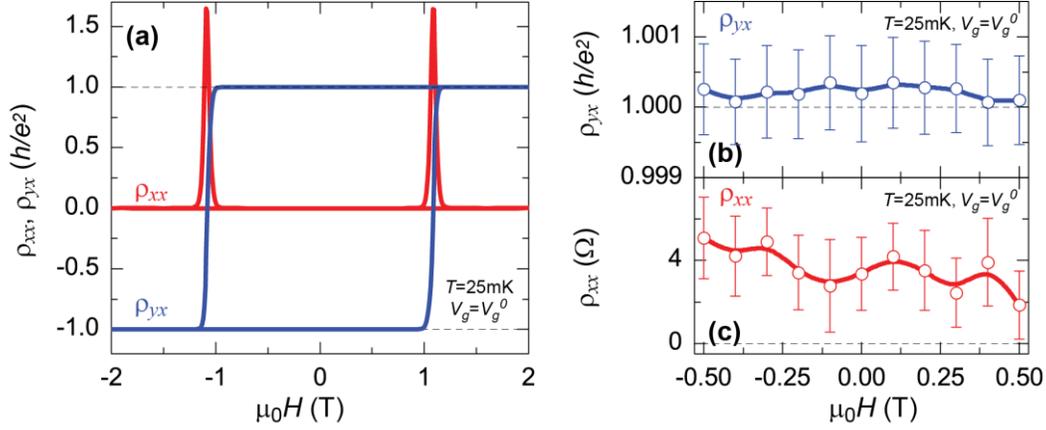

**Fig. 24| The QAHE in a 4QL (Bi$_{0.29}$Sb$_{0.71}$)$_{1.89}$V$_{0.11}$Te$_3$ film measured at 25 mK.** (a) Magnetic field dependence of the longitudinal resistance $\rho_{xx}$ (red) and the Hall resistance $\rho_{yx}$ (blue) at charge neutral point $V_g = V_g^0$. (b, c) Expanded $\rho_{yx}$ (b) and $\rho_{xx}$ (c) at low magnetic field. $\rho_{yx}$ at zero magnetic field exhibits a value of $1.00019 \pm 0.00069$ $h/e^2$, while $\rho_{xx}$ at zero magnetic field is only $\sim 0.00013 \pm 0.00007$ $h/e^2$ ($\sim 3.35 \pm 1.76$ $\Omega$). Adapted from Ref. [7].

The temperature dependence of QAH behavior at the charge natural point $V_g^0$ after magnetic training is shown in Fig. 25a. Below $T = 5$ K, $\rho_{yx}$ (same as $\rho_{yx}(0)$) shows a continuous increase and $\rho_{xx}$ (same as $\rho_{xx}(0)$) shows a continuous decrease at lower temperature, indicating that the QAH state could survive as high as ~5 K, compared to the Cr-doped system which is only attainable at 30 mK. Fig. 25b displays the temperature dependence of the anomalous Hall angle $\alpha$. The $\alpha$ grows rapidly with decreasing temperature, and asymptotically approaches the ideal value 90 ° at the lowest temperature ~25 mK, which is consistent with previous Hall results.

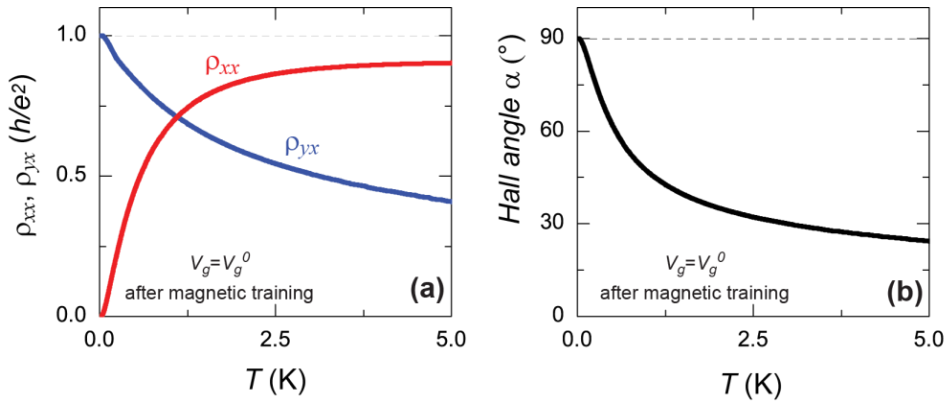

**Fig. 25 | The temperature dependence of QAH behavior in a 4QL (Bi$_{0.29}$Sb$_{0.71}$)$_{1.89}$V$_{0.11}$Te$_3$ film.** (a) Temperature dependence of the longitudinal resistance $\rho_{xx}$ (red) and Hall resistance $\rho_{yx}$ (blue) of sample a 4QL (Bi$_{0.29}$Sb$_{0.71}$)$_{1.89}$V$_{0.11}$Te$_3$ after magnetic training. (b) Temperature dependence of the anomalous Hall angle $\alpha$ also shows is the ideal value (90°, dashed line). Adapted from Ref. [7].

Fig. 26a shows the temperature-dependence of $\rho_{xx}$ and $\rho_{yx}$ for another 4QL (Bi$_{0.29}$Sb$_{0.71}$)$_{1.89}$V$_{0.11}$Te$_3$ film (same thickness and same composition as above), using a bottom gate bias $V_g$=0 and zero-field cooling. $\rho_{xx}$ shows semiconducting behavior from room temperature down to ~2.5 K due to the lower thermally-excited carrier density as temperature is lowered, followed by a sharp drop, signaling a phase transition toward the QAH state. $\rho_{yx}$ remains constant from room temperature down to ~23 K where it experiences an upturn, indicating a $T_C$~23 K FM state due to the existence of the AHE. $\rho_{yx}$ continues to increase and $\rho_{xx}$ continues to decrease with decreasing temperature. $\rho_{yx}$ approaches to ~0.99 $h/e^2$ and $\rho_{xx}$ reaches ~0.30 $h/e^2$ at ~80 mK. This behavior signifies the film spontaneously progressing toward the QAH state. The net magnetization of virgin state in the film, *i.e.* not induced by any external field, and the reduced thermal activation at lower temperature drives the film from the diffusive transport regime toward the conducting QAH state. Fig. 26a inset shows the low-temperature transport data at $V_g$=0 below 3 K at magnetic fields μ$_0$H=0 and 2 T. This spontaneous drop of $\rho_{xx}$ with V-doping is absent in Cr-doped TIs. In Cr-doped TI films, unless the magnetic domains are aligned with an external applied perpendicular magnetic field, $\rho_{xx}$ will increase when temperature is lowered and no QAH state can be achieved even at very low temperatures[71]. On the other hand, for the V-doped TI film, even a strong field of 2T did not change the qualitative behavior of $\rho_{xx}$ and $\rho_{yx}$. This clearly demonstrates that the macroscopic net magnetization of the virgin state in V-doped TIs is spontaneous and highly developed. The fact that the virgin $\rho_{yx}$ and $\rho_{xx}$ curves overlap with the hysteresis loop (Fig. 26b) as well as the butterfly structure (Fig. 26c), also confirms the spontaneous self-magnetization of V-doped TI samples in the QAH regime.

## 5.4 The relationship between the impurity bands and QAHE

The QAH state observed in V-doped TIs is unexpected from the view of first-principle calculations, which showed a metallic FM state in V-doped TI due to the presence of *d*-orbital impurity bands at the Fermi level [19]. A possible explanation why the *d*-orbital impurity bands do not contribute to the charge transport and results in an insulating FM state is that this impurity bands may be localized at low temperature due to Anderson localization. The localization physics in the QAH regime has been studied numerically [81], which confirms that the QAH state can still be achieved in a metallic system by introducing disorder to localize bulk carriers, which is consistent with our experimental observation of the role of localization in impurity bands.

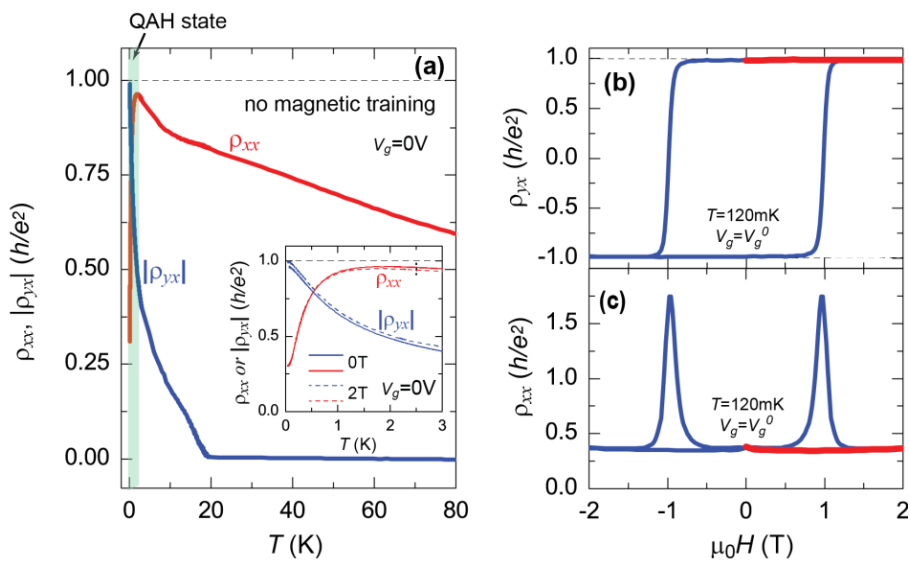

**Fig. 26| The self-driven QAH state in a different 4QL (Bi$_{0.29}$Sb$_{0.71}$)$_{1.89}$V$_{0.11}$Te$_3$ film.** (a) Temperature dependence of longitudinal resistance $\rho_{xx}$ (red) and Hall resistance $|\rho_{yx}|$ (blue) without magnetic training. Inset: *T*-dependent $\rho_{xx}$ (red) and $|\rho_{yx}|$ (blue) under magnetic field $\mu_0 H$=0 (solid) and 2T (dashed). The opposite *T*-dependent behaviors of $\rho_{xx}$ and $/\rho_{yx}/$ reveal the QAH state at low *T*. (b, c) The magnetic field

dependent $\rho_{yx}$ (b) and $\rho_{xx}$ (c) measured at 120 mK. The red curves in (b) and (c) were obtained by first cooling the sample under zero-field and then measured from zero to higher field. Adapted from Ref. [7].

This physical picture also provides a natural explanation of high $T_C$ observed in V-doped Sb$_2$Te$_3$ compared to that in Cr-doped Sb$_2$Te$_3$. The van Velck mechanism depends on the inverted band structure, and is similar for these two systems since their parent materials are identical. The impurity bands can mediate a double-exchange interaction between magnetic moments, which is well known for some DMSs, particularly in the insulating regime [82]. Thus, we expect that the double-exchange interaction provides additional channels for FM order, leading to the enhancement of the $T_C$ in V-doped Sb$_2$Te$_3$. This explanation is further supported by the fact that the gate-voltage dependence of $H_c$ of V-doped film is much larger than that of Cr-doped system [7]. This scenario paves a feasible path to enhance the observation temperature of the QAHE.

### 5.5 Zero-field dissipationless chiral edge transport in QAHE

In Section 5.3, we have shown that the zero-field quantized Hall plateau was accompanied by a negligible $\rho_{xx}$. However, this observation has yet to be identified directly that the dissipationless transport that occurs via chiral edge channels, which is a feature of the QAHE. In particular, the physical origin of dissipation had yet to be clarified. The identification of dissipative channels is essential in attempts to increase the critical temperature of the QAH state, which is almost two orders of magnitude lower than the $T_C$ in magnetically doped (Bi,Sb)$_2$Te$_3$ [6,7,70,71]. To address these issues, we turn to nonlocal transport measurements, which provided definitive experimental evidence for the existence of chiral edge channels in the QH state at high magnetic fields [2] and also for helical edge channels in the QSH state at zero magnetic field [83]. Thus we combine two-, three- and four-terminal local and nonlocal measurements to extract information on different conducting channels and directly reveal the zero-field dissipationless nature of chiral edge modes and the onset of dissipative channels for the QAH state in 4QL V-doped $(Bi,Sb)_2Te_3$.

Starting from local transport measurements for different terminals, Fig. 27a shows the $\mu_0H$ dependence of the two-terminal resistances measured at $T$=25 mK, using a bottom gate bias $V_g = V_g^0 = +7$ V to reach the charge neutral point. The two-terminal resistances $\rho_{12,12}$, $\rho_{13,13}$ and $\rho_{14,14}$ show exactly $h/e^2$ quantized values when the sample is well magnetized. (Note that the first two subscripts of the resistivity refer to the current leads and the last two to the voltage leads.) However, local two-terminal resistance measurement cannot reveal the chirality of the edge transport explicitly.

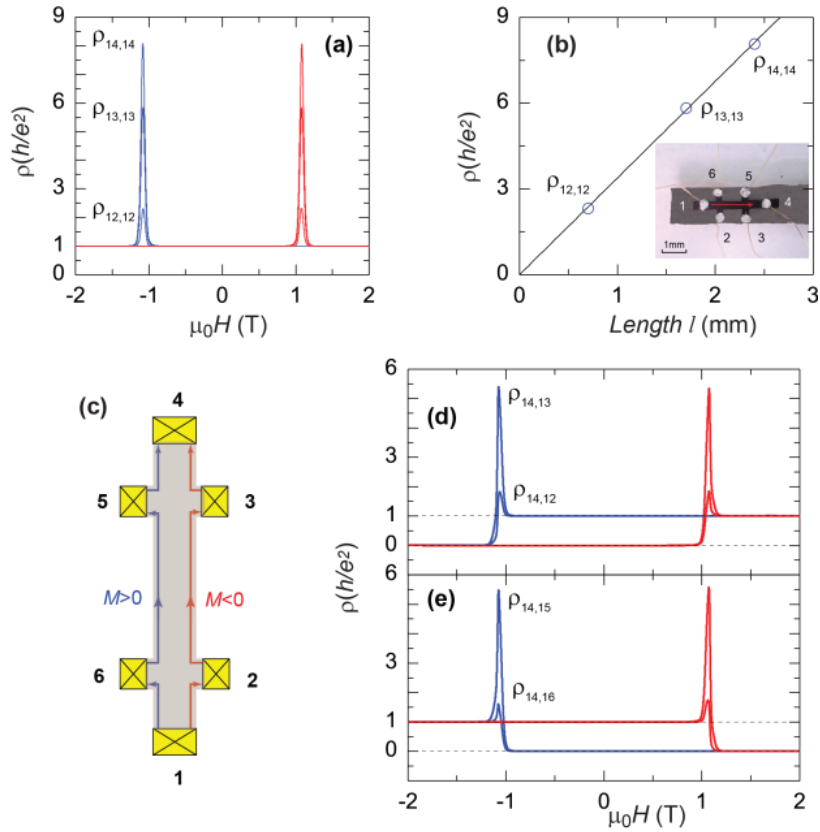

**Fig.27| Local two-terminal and three-terminal measurements in the QAH regime.** (a) Magnetic field ($\mu_0H$) dependence of two-terminal local resistances $\rho_{12,12}$, $\rho_{13,13}$ and $\rho_{14,14}$. (b) The peak value near the

coercive field ($H_c$) of two-terminal resistances $\rho_{12,12}$, $\rho_{13,13}$ and $\rho_{14,14}$ as a function of the spacing length $l$(mm) between the voltage electrodes. The inset photograph shows the Hall bridge device. (b) Schematic layout of the device applicable for panels (d) and (e) and also for $\rho_{14,14}$ in (a). The current flows from 1 to 4. The red and blue lines indicates the chiral edge current for magnetization into ($M<0$) and out of the plane ($M>0$), respectively. (d, e) $\mu_0 H$ dependence of three-terminal resistances $\rho_{14,13}$, $\rho_{14,12}$ (d) and $\rho_{14,15}$, $\rho_{14,16}$ (e). Adapted from Ref. [78].

To probe the detailed flow of the edge channels, we have performed local three-terminal resistance measurements at $T=25$ mK with $V_g=V_g^{~0}$ (Figs. 27d and 27e). The resistances $\rho_{14,13}$ and $\rho_{14,12}$ (Fig. 27d) display similar loops that are asymmetric. At $\mu_0 H=0$, the $\rho_{14,13}$ and $\rho_{14,12}$ values depend on the magnetization $M$ orientation, with a value of $h/e^2$ for $M>0$ and 0 for $M<0$, respectively. The resistances $\rho_{14,16}$ and $\rho_{14,15}$ display loops that are mirror symmetric to $\rho_{14,12}$ and $\rho_{14,13}$ by $\mu_0 H \to -\mu_0 H$ (Fig. 27e). At $\mu_0 H=0$, $\rho_{14,16}$ and $\rho_{14,15}$ are 0 for $M>0$ and $h/e^2$ for $M<0$. The asymmetric loops of three-terminal resistance and the relation between $\rho_{14,16}$ ($\rho_{14,15}$) and $\rho_{14,12}$ ($\rho_{14,13}$) are direct manifestations of the chirality of edge transport, which can be understood from the Landauer-Buttiker formalism [84,85]. In the QAH regime for $M<0$, since chiral edge modes propagate anticlockwise ($1 \to 2 \to 3 \ldots$), the transmission coefficients, denoted as $T_{ij}$ from electrode $j$ to $i$, are non-zero only for $T_{21}=T_{32}=T_{43}=1$. When current flows from electrode 1 to 4 (Fig. 27c), the voltage distributions are $V_2=V_3=V_1=(h/e^2)I$ and $V_6=V_5=V_4=0$, where $V_i$ denotes the voltage at the electrode $i$. Thus, the corresponding resistance is $\rho_{14,13}=\rho_{14,12}=0$ and $\rho_{14,14}=\rho_{14,16}=\rho_{14,15}=h/e^2$. Likewise for $M>0$, chiral edge modes travel clockwise and the nonzero transmission matrix elements are $T_{61}=T_{56}=T_{45}=1$, leading to the resistance $\rho_{14,14}=\rho_{14,13}=\rho_{14,12}=h/e^2$ and $\rho_{14,16}=\rho_{14,15}=0$. The electrical potential distributions for $M<0$ and $M>0$ (see Fig. 28) can also be calculated using the conformal-mapping technique [86]. Hence at $\mu_0 H=0$, the two-terminal resistance $\rho_{14,14}$ is always quantized at $h/e^2$ regardless of the $M$ orientation. In contrast, the three-terminal local resistances $\rho_{14,13}$ (or $\rho_{14,12}$) and $\rho_{14,15}$ (or $\rho_{14,16}$) depend on the $M$ orientation.

Both two- and three-terminal resistances deviate from the quantized value near the coercive fields ($H_c$) giving rise to two sharp resistance peaks (Figs. 27a, d and e). This corresponds to a plateau-to-plateau transition (PPT) region. The heights of the resistance peaks for different terminal measurements are different. We plot the highest resistance values of $\rho_{12,12}$, $\rho_{13,13}$ and $\rho_{14,14}$ as a function of the distance between these terminals (Fig. 27b), and found a linear dependence between resistance and distance. This suggests that in the PPT region, the transport occurs through the bulk of the system. In contrast, the two-terminal resistance of any pair of electrodes is always $h/e^2$ and independent of the length in the QAH regime, which is further consolidated by various quantized rational numbers for different interconnections among the electrodes at the periphery of the six-terminal Hall bridge (Fig. 29). This $h/e^2$ quantized two-terminal resistance is similar to that in the QHE of 2DEGs [87,88], with the important distinction that the ballistic edge channels of the QAH state survive without an external magnetic field. Furthermore, the ballistic transport with $h/e^2$ quantized resistance indicates the propagation of single spin species, in contrast to the QH state for which the spin polarization of the edge channels requires Zeeman coupling to an externally (usually large) applied magnetic field [1,2].

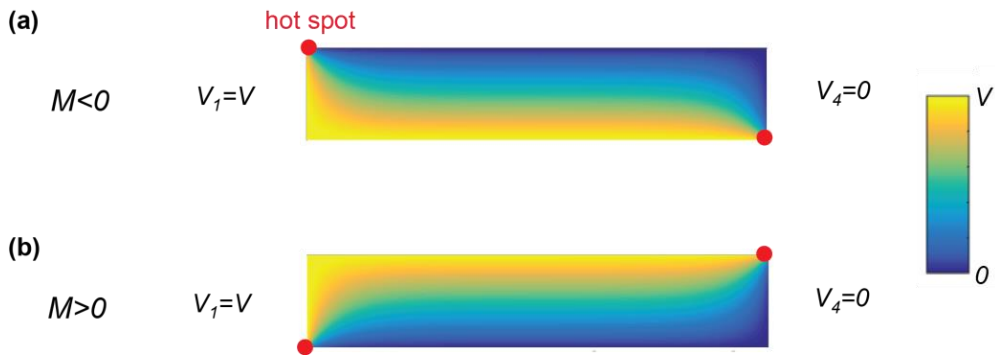

**Fig.28| The electric potential distributions of a six terminal Hall bridge in the QAH regime for *M*<0 (a) and *M*>0 (b).** The calculation assumes the tangent of Hall angle tan$\alpha$=7000, the length and the width of the samples are *L* and *W*, and *L*=5*W*. Adapted from Ref. [78].

By comparing local with non-local measurements, we can also reveal distinct behaviors for ballistic chiral edge transport and diffusive bulk transport. In Fig. 30a, the current was passed through electrode 3 to 1, while the three-terminal local and nonlocal resistances $\rho_{31,32}$, $\rho_{31,34}$, $\rho_{31,35}$, and $\rho_{31,36}$ were measured at *T*=25 mK with $V_g=V_g^0$. The $\mu_0H$ dependence of $\rho_{31,32}$ is also symmetric to those of $\rho_{31,34}$, $\rho_{31,35}$, and $\rho_{31,36}$ (Fig. 30b) by $\mu_0H \rightarrow -\mu_0H$, thanks to the chirality of the edge current. An interesting feature is that $\rho_{31,32}$ and $\rho_{31,36}$ show prominent peaks in the PPT region with a value of a multiple of $h/e^2$, while $\rho_{31,34}$ and $\rho_{31,35}$ vary between $h/e^2$ and 0 smoothly without any peaks. The peaks in $\rho_{31,32}$ and $\rho_{31,36}$ are induced by contributions from the local diffusive longitudinal resistance through the bulk. Since no local $\rho_{xx}$ were picked up, no peaks appeared in $\rho_{31,34}$ and $\rho_{31,35}$.

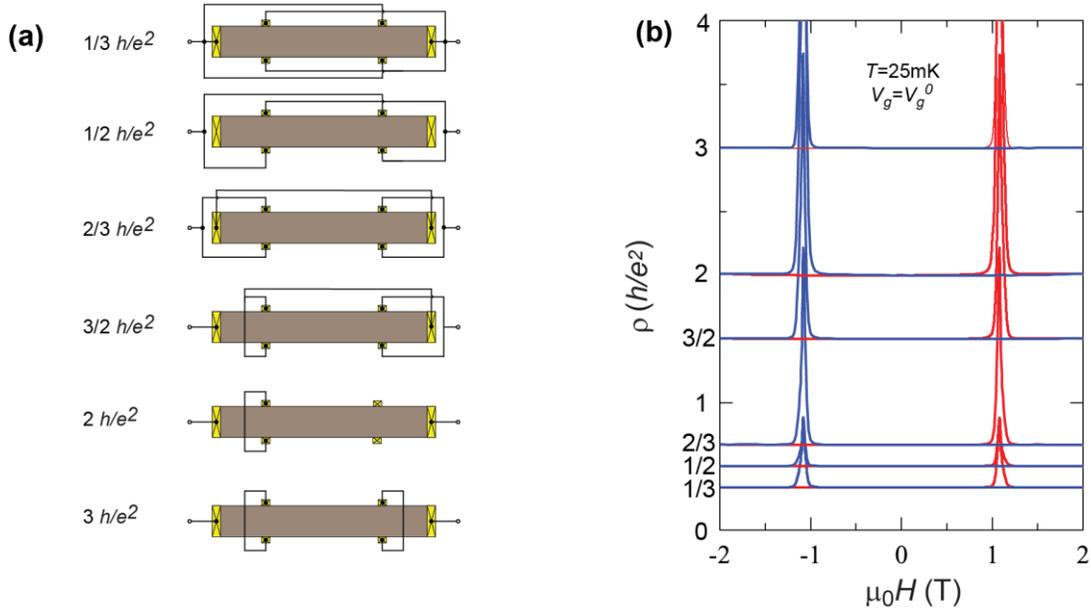

**Fig. 29| Quantized two-terminal resistance in multiply-connected perimeters.** (a) The perimeter interconnection configurations of the six-terminal Hall bridge. The quantized values of two-terminal resistance are indicated on the left of the configurations. (b) $\mu_0 H$ dependence of two-terminal resistance for each configuration in (a). Adapted from Ref. [78].

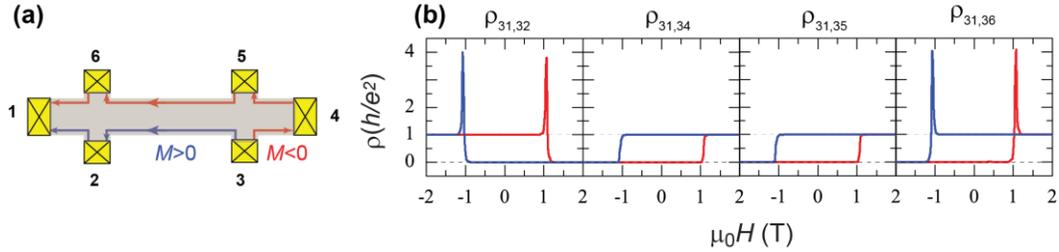

**Fig. 30| Local and nonlocal three-terminal measurements in the QAH regime.** (a) Chiral edge conduction channels when the current flows from 3 to 1. (b) $\mu_0 H$ dependence of local and nonlocal three-terminal resistances $\rho_{31,32}$, $\rho_{31,34}$, $\rho_{31,35}$, and $\rho_{31,36}$ measured at $T$=25 mK with $V_g = V_g^0$. Adapted from Ref. [78] .

## 5.6 The nature of dissipation in the QAHE

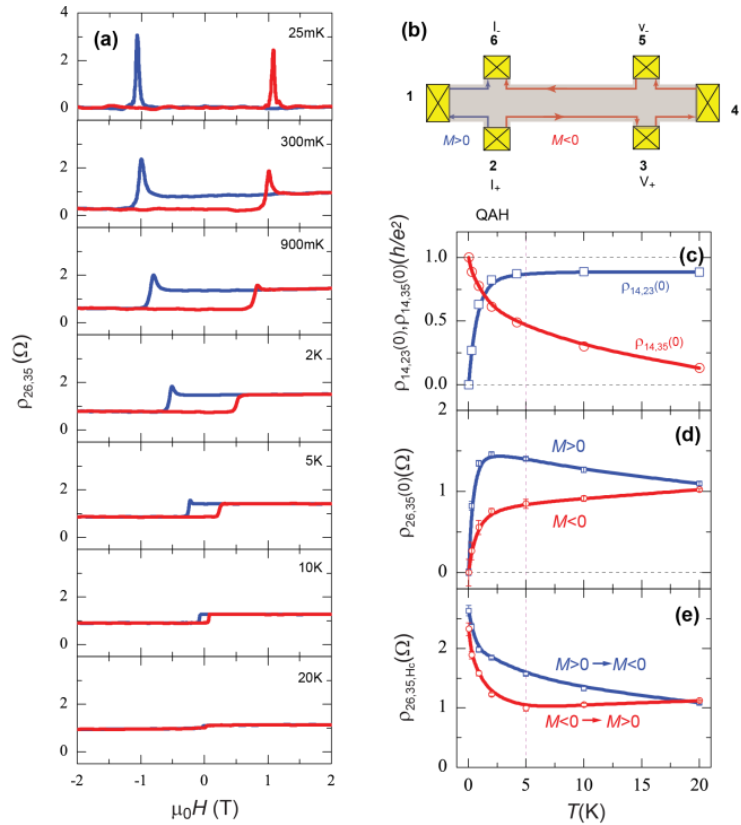

**Fig. 31|Temperature dependence of chiral edge transport in the QAH regime.** (a) $\mu_0 H$ dependence of the nonlocal resistance $\rho_{26,35}$ measured at $V_g = V_g^0$ from 25 mK to 20 K. (b) Chiral edge conduction channels when the current flows from 2 to 6, and the nonlocal voltage is measured between 3 and 5. (c) Temperature dependence of the zero-field longitudinal sheet resistance $\rho_{14,23}(0)$ (blue) and Hall resistance $\rho_{14,35}(0)$ (red). (d) Temperature dependence of zero-field nonlocal signal $\rho_{26,35}(0)$ for $M<0$ (red) and $M>0$ (blue), respectively. (e) Temperature dependence of nonlocal resistance $\rho_{26,35}$ peaks in the PPT as seen in (a) going between two magnetization orientations. Adapted from Ref. [78].

As described above, we have shown that by carefully tuning the $V_g$, we can achieve, at extremely low temperatures, purely dissipationless chiral edge transport, and the bulk transport only occurs in the PPT region. Next we will explore the origin of dissipation at zero magnetic field when the temperature is increased. We focus on the four-terminal nonlocal measurement configuration, where electrodes 2 and 6 were designated as the current electrodes while electrodes 3 and 5 were used as the voltage probes (Fig. 31b). In Figs. 31a and 32h, at the lowest temperature $T=25$ mK, $\rho_{26,35}=0$ (in the non-PPT region) regardless of the direction of $M$, consistent with the picture of a pure chiral edge transport. With increased temperature, $\rho_{26,35}$ exhibits a hysteresis loop with a decreased $H_c$ (Fig. 31a). The observation of resistance hysteresis indicates the emergence of dissipative channels in addition to chiral edge modes. Fig. 31d shows the zero-field non-local resistance as a function of temperature, which increase rapidly with temperature up to 2K, accompanied by an increase of longitudinal resistance $\rho_{14,23}$ and a decrease of Hall resistance $\rho_{14,35}$ (Fig. 31c). At the same time, the resistance peaks in the PPT region decay rapidly (Fig. 31e). The different temperature dependences between zero field nonlocal signals and the resistance peak in the PPT region suggest that they should have different origins. In Fig. 27b, we have confirmed that the resistance peak in the PPT region follows a linear dependence on distance, thus originating from the bulk carriers. This linearity and the different temperature dependences indicate that the bulk carriers are not responsible for nonlocal signals.

A possible explanation comes from the existence of nonchiral edge channels, which have been invoked previously to explain a similar hysteresis loop of nonlocal voltage in Cr-doped $(Bi,Sb)_2Te_3$ [76]. In that experiment, however, the hysteresis loop was observed at the lowest temperature, indicating the existence of gapless quasihelical edge modes [89]. For our system, the observation of zero longitudinal and nonlocal resistances at the lowest reachable temperature (with $V_g = V_g^0$) rules out gapless modes. Nevertheless, gapped nonchiral edge modes are possible, and, as we argue below, plausible. These nonchiral edge modes originate from two dimensional Dirac surface states on the side surfaces, which are quantized into one dimensional edge modes due to the confinement of finite thickness [89,90].

Our physical picture that the hysteresis loop of nonlocal signals arises from the coexistence of chiral and nonchiral edge modes finds strong support from the $V_g$ dependence of local and nonlocal measurements. Figs. 32a to 32e show the longitudinal sheet resistance $\rho_{14,23}$ and Hall resistance $\rho_{14,35}$ at different $V_g$, and the corresponding nonlocal resistance $\rho_{26,35}$ is shown in Figs. 32f to 32j. A pronounced asymmetry between $V_g > V_g^0$ and $V_g < V_g^0$ is observed. At $T$=25 mK with $V_g = V_g^0$ when the Fermi energy is in the excitation gap, $\rho_{14,35}$ is fully quantized and $\rho_{14,23}$ simultaneously vanishes. The value of nonlocal resistance $\rho_{26,35}$ is always 0 except at the PPT region (see Fig. 32h). For $V_g > V_g^0$, $\rho_{14,23}$ and $\rho_{14,35}$ have almost no significant change (Figs. 32d and 32e), while a hysteresis loop appears for the non-local resistance $\rho_{26,35}$ (Figs. 32i and 32j). In contrast, for $V_g < V_g^0$, a huge $\rho_{14,23}$ is observed while the nonlocal resistance $\rho_{26,35}$ is always close to zero (Figs. 32f and 32g). The variation of $\rho_{14,23}$ and $\rho_{14,35}$, as well as the nonlocal resistance $\rho_{26,35}$, as a function of $V_g$ is summarized in Figs. 32k and 32l. The ratio between $\rho_{26,35}$ and $\rho_{14,23}$ is approximately $10^{-2}$ for $M>0$ in the $V_g>V_g{}^0$ regime, significantly larger than its value $10^{-5}$ in the $V_g<V_g{}^0$ regime, as shown in Fig. 32m.

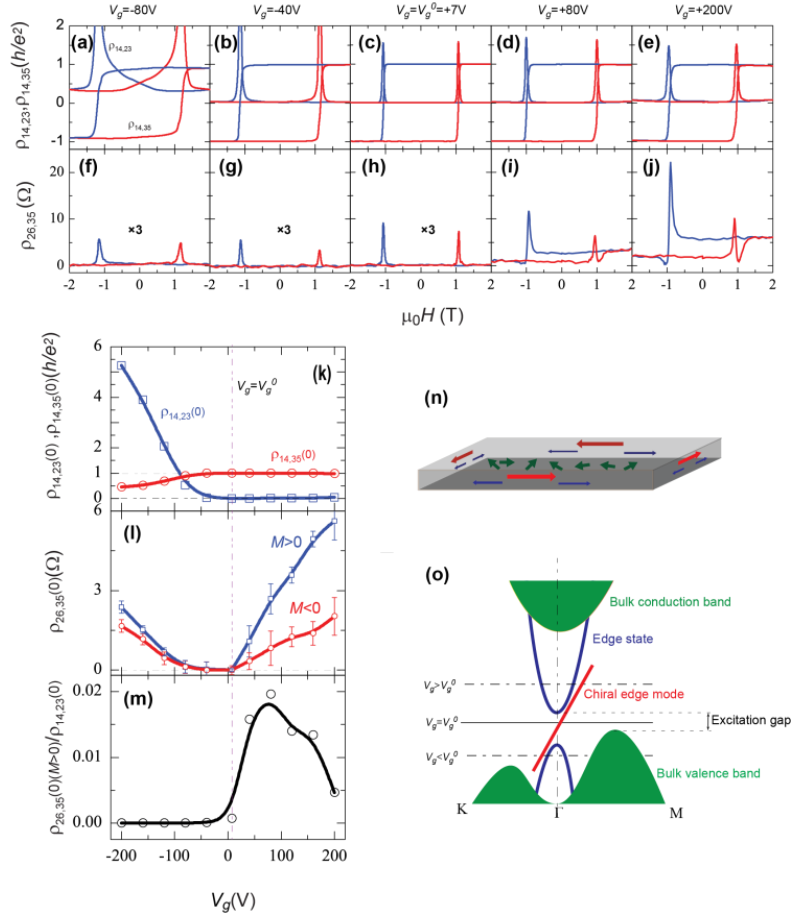

**Fig. 32| Gate bias dependence of chiral edge transport in the QAH regime.** (a to j) $\mu_0H$ dependence of longitudinal resistance $\rho_{14,23}$ and Hall resistance $\rho_{14,35}$ (a to e), as well as the nonlocal signal $\rho_{26,35}$ (f to j) at various $V_g$s. (k) $V_g$ dependence of the zero-field longitudinal sheet resistance $\rho_{14,23}(0)$ (blue) and Hall resistance $\rho_{14,35}(0)$ (red). (l) $V_g$ dependence of zero-field nonlocal signal $\rho_{26,35}(0)$ for $M<0$ (red) and $M>0$ (blue), respectively. (m) $V_g$ dependence of the ratio between nonlocal resistance $\rho_{26,35}(0)$ for $M>0$ and longitudinal sheet resistance $\rho_{14,23}(0)$. Note that the nonlocal resistance $\rho_{26,35}(0)$ for $M>0$ indicates that the dissipation channels are in the voltage probe side. (n) The schematic diagrams identify three channels in the sample: dissipationless edge channels (red arrows), dissipative edge channels (blue arrows) and dissipative bulk channels (green arrows). (o) The schematic band dispersions of the sample. The horizontal dash-dotted line indicates the Fermi level position for $V_g>V_g{}^0$ and $V_g<V_g{}^0$, respectively. The

excitation gap is defined as the energy difference between the bottom of the non-chiral edge mode and the maximum of bulk valence band as indicated. Adapted from Ref. [78].

The asymmetry between $V_g > V_g^0$ and $V_g < V_g^0$ can be understood from the detailed band structure of $(Bi,Sb)_2Te_3$ and the positioning of the gap relative to the valence and conduction bands. From previous ARPES measurements [66,91] and first principles calculations [24,45], it is known that the surface Dirac cones are far away from the bulk conduction band bottom and quite close to the valence band. For our magnetic TI system, the energy spectrum is schematically shown in Fig. 32o. The green part represents the 2D bulk bands of the thin film, which originates from both the 3D bulk bands and 2D surface bands of top and bottom surfaces. The gap of 2D bulk bands should be determined by the exchange coupling between surface states and magnetization $M$. Within the 2D bulk gap, there are two types of 1D edge modes: the chiral modes and the nonchiral edge modes originating from surface states of side surfaces as discussed earlier [89,90]. The pure dissipationless chiral edge transport only occurs when the Fermi energy is tuned into the mini-gap of the nonchiral edge modes, which is induced by the confinement effect of the side surface and lies close to the maximum of the valence band. For $V_g > V_g^0$, the Fermi energy first cuts through the nonchiral edge modes, leading to a hysteresis loop of nonlocal resistance due to the coexistence of two types of edge modes. In contrast, for $V_g < V_g^0$, the Fermi energy will first encounter the top of 2D (bulk) valence bands, resulting in a large $\rho_{14,23}$ and an insignificant nonlocal resistance $\rho_{26,35}$. It is known that the classical longitudinal transport can also contribute to nonlocal effects, for which the ratio between nonlocal resistance $\rho_{26,35}$ and $\rho_{14,23}$ can be estimated as $\frac{\rho_{26,35}^{cl}}{\rho_{14,23}} \approx e^{-\frac{\pi L}{W}}$, where $L$ and $W$ are the length and width of the sample, respectively [89]. In our case, $\frac{L}{W} \sim 3$, so this ratio is estimated $\sim 8 \times 10^{-5}$, which can explain the observed nonlocal resistance ratio $(10^{-5})$ in the $V_g < V_g^0$ regime, but not that $(10^{-2})$ in the $V_g > V_g^0$ regime.

Therefore, nonlocal signals as well as the hysteresis loop for $V_g > V_g^0$ should be dominated by the mechanism of the coexistence of chiral and nonchiral edge modes.

The above physical picture is also consistent the temperature dependence of longitudinal and nonlocal resistances in this system. With increasing temperature, $\rho_{14,23}$ increases rapidly (Fig. 32c), indicating the existence of bulk carriers. At the same time, the observation of hysteresis loop suggests that nonchiral edge modes should also appear. According to the band dispersion in Fig. 32o, we speculate that finite temperature excites electrons from 2D valence bands to 1D nonchiral edge channels, so that both 2D bulk holes and 1D nonchiral edge electrons coexist in the system (Fig 32n). The excitation gap, as indicated in Fig. 32o, is estimated as 50 μeV by fitting temperature dependence of $\rho_{14,23}$, which is consistent with the theoretical prediction [78]. This excitation gap is expected to be much smaller than the 2D bulk gap due to magnetization and is consistent with the low critical temperature for the QAH effect. Besides these two kinds of dissipative channels, one should note that other states, such as acceptor or donor states due to impurities, could also exist in the system and cause dissipation.

Although QAHE in the above Cr- and V-doped TI system exhibits similar properties as the QHE, the underlying physical mechanism is distinct. QHE results from Landau levels formed in a 2DEG under a strong external magnetic field. QAHE, however, occurs at zero external magnetic field without Landau levels but requires intrinsic magnetization. Ideally, the disorder inside the film cannot only break the QAH state, but also benefit to localize the dissipative channels. Up to now, in all QAHE experiments [6,7,70,71], the motilities of the films are below ~1000cm$^2$/Vs, orders of magnitude smaller than that found in semiconductor hetero-structures with 2DEG that reached the QH state. On the other hand, the fluctuation of the electrical band (Fig. 32o) induced by the disorder will drown the ferromagnetic exchange gap, killing the QAH state. Therefore, only the samples with appropriate doping level can show QAH state. Further doping-level dependence QAHE investigations are highly desired.

## 6. Conclusions and Outlook

We have reviewed the members of the Hall family, which contain the ordinary, the spin and the anomalous Hall effects, with both nonquantized and quantized versions. Based on this, we focus on the route leading to the recent experimental observation of the QAHE in Cr- and V-doped $(Bi,Sb)_2Te_3$ TIs, and the current status of the research of the QAHE. The experimental observation of the QAHE concludes a decades-long search for the QHE without Landau levels [8]. Moreover, the exactly zero-field dissipationless chiral edge transport property in the QAHE of V-doped TI system has opened up the possibility in electronic devices with low-power consumption. Hereinafter, we propose a few prospective future directions:

(a) The elevation of QAHE observation temperature. At the current stage, the major challenge preventing device applications on the QAHE is the ultralow temperature (<100 mK) needed to reach the quantized plateau and pure dissipationless chiral edge mode. The identification of dissipative channels in QAH regime suggests ways to increase the observing temperature of the QAHE in Cr- or V-doped TI systems. First and foremost, even thinner TI films (<4QL) can be utilized to reduce the number of nonchiral edge channels induced by the finite thickness, and increase the gap between nonchiral edge modes and bulk valence bands. Secondly, the concentration of Bi dopants should be reduced to lower the valence band maximum, pushing all edge modes above valance band. Thirdly, the interfacial magnetic order should be enhanced. Recently, using a hybrid structure composed of both doping and proximity effect, we have observed an enhancement of proximity magnetic order using polarized beam neutron reflectometry [92]. Last but not least, other FM insulators with high $T_C$, such as yttrium iron garnet (YIG) ($T_C$~550 K) [93] in addition to EuS (~16.5 K) [59] and EuO (Tc~77 K), can be applied as proximity layer. The high $T_C$ of FM insulators is very beneficial for realizing the QAH state at higher temperature. Up to now, the main point in this direction is to enhance the exchange coupling between TI and FM insulator by improving the interface quality between them. One effective solution is that both the TI and FM insulator should be grown in the same chamber in order to avoid the oxidation and water adsorption when the sample is taken out of the chamber. More studies are deserved to explore the high observing temperature of QAHE.

(b) QAHE with higher Chern numbers. The QAHE with higher Chern numbers has been theoretically proposed from more band inversions among subbands, which are induced by larger magnetic exchange fields [94,95]. Now, the challenge of obtaining these higher Chern number QAHE is that when the exchange field splitting becomes greater than the bulk band gap, the system usually becomes completely metallic before the higher plateaus emerge. Another interesting proposal is based on materials with intrinsic multiple band inversions, *e.g.* magnetically-doped topological crystalline insulators (TCI), such as (Pb,Sn)(Se,Te), or TCI in proximity to FM insulators, in both of which the van Vleck mechanism also plays an important role. The QAHE with any Chern number between ±4 could be realized in thin films with proper magnetic doping concentrations, structural distortion and thickness [96,97].

(c) Fractional QAHE. It has been known for decades that the strong correlation of 2DEG under high magnetic fields results in fractionally-charged and fractional statistics of quasiparticles, called fractional QHE [98]. Recently, a few groups including Wen, Mudry and Das Sarma [99-101] have demonstrated the possibility of FQHE in the absence of an external magnetic field. The FQHE without Landau levels inspires us to consider the possibilities of fractional QAHE [102], as Haldane's model of QHE inspired the QAHE. However, the suitable choice of material system would be crucial. The DFT calculations that pioneered the research of TI and QAHE in realistic materials have limited power due to the possible strong correlation nature of the quasiparticle in the fractional QAHE.

(d) Universal scaling behavior in QAHE. In the QH regime, the PPT reflects the localization-delocalization-localization transition in adjacent Landau levels [103,104]. While in the QAH regime, the magnetic domains of the film can be considered as a single domain with down or up magnetization, corresponding to the quantized $\rho_{yx}$ plateaus of $-h/e^2$ or $h/e^2$. The PPT of the QAHE is induced by the magnetization reversal process. The PPT in the QAH regime is predicted to have a similar scaling behavior as that in the QH regime [75]. Until now, it is uncertain whether the PPT in the QAHE belongs to the same universality class as that in the QH regime. In this regards, a direct and systematic investigation aimed at the scaling behavior in the QAHE is highly desirable.

(e) Device integration and mass production. To bring QAHE into real-world applications, in addition to a high working temperature, we must bring technologies suitable for low-cost sample fabrication and device integration into the semiconductor industry. Up to now, QAH samples only can be fabricated in the lab via MBE [6,7,70,71,78], thus, more general and convenient methods, e.g. pulsed laser deposition (PLD) and sputtering, or other conceptually new methods, are highly desired.

(f) Realization of chiral topological superconductor. The chiral topological superconductor has been predicted to exist in a hybrid device of a QH or QAH insulator near the plateau transition in proximity to a fully gaped s-wave superconductor [105]. Compared to QH systems, the QAH state can be realized without external magnetic field, making the realization of superconductor proximity effect much easier. The chiral topological superconductor in a hybrid structure of the QAH insulator with superconductor only holds for an odd number of chiral Majorana zero modes, which can be detected by tunneling junctions. The QAHE with s-wave superconductors in proximity can thus serve as building blocks for topological quantum computing, where the entangled Majorana zero modes behave as one qubit.

## Acknowledgements

We are grateful to Q. K. Xue, K. He, Y. Wang, L. Lv, D. Heiman, S. C. Zhang, J. S. Moodera, C. X. Liu, H. Zhang, P. Wei, W. Zhao, J. K. Jain, M. H. W. Chan, J. Li and Y. Zhu for fruitful collaborations and helpful discussions. This work is supported by the STC Center for Integrated Quantum Materials under NSF grant DMR-1231319.